\documentclass{jpsj2}
\title{Local-Ansatz Approach with Momentum Dependent Variational
Parameters to Correlated Electron Systems}

\author{Yoshiro \textsc{Kakehashi}\thanks{E-mail address:
yok@sci.u-ryukyu.ac.jp, to be published in Phys. Soc. Jpn. {\bf 77} No.11 
(2008)},
Takafumi \textsc{Shimabukuro}, and Chitoshi \textsc{Yasuda}
}

\inst{
Department of Physics and Earth Sciences,
Faculty of Science, University of the Ryukyus, \\
1 Senbaru, Nishihara, Okinawa, 903-0213, Japan
}

\abst{
A new wavefunction which improves the Gutzwiller-type local ansatz
method has been proposed to describe the correlated electron system.  
The ground-state energy, double occupation number, momentum distribution 
function, and quasiparticle weight have been calculated for the
half-filled band Hubbard model in infinite dimensions.  It is shown that
the new wavefunction improves the local-ansatz approach (LA) proposed by 
Stollhoff and Fulde.  Especially, calculated momentum distribution
functions show a reasonable momentum dependence.  
The result qualitatively differs from those obtained by the LA 
and the Gutzwiller wavefunction.  Furthermore,
the present approach combined with the projection operator method CPA 
is shown to describe quantitatively the excitation spectra in
the insulator regime as well as the critical Coulomb interactions for a
gap formation in infinite dimensions.
}

\kword{variational method, electron correlations, Gutzwiller
wavefunction, local ansatz, Hubbard model, excitation spectra, critical
Coulomb interaction, infinite dimensions}

\begin{document}
\maketitle

\section{Introduction}
Variational method has been a useful tool to investigate the
ground-state properties of correlated electrons from molecules to solids
over half a century.  In the method, a minimum basis set to describe
correlated electrons is
constructed by applying one-particle, two-particle,
and higher-order particle operators onto the Hartree-Fock wavefunction,
and their amplitudes are chosen to be best on the basis of
the variational principle.  Among various trial wavefunctions,
the Gutzwiller wavefunction is one of the simplest and popular
wavefunctions in solids.  It was introduced by Gutzwiller to clarify the
role of electron correlations in metallic
ferromagnetism~\cite{gutz63,gutz64}.   
The idea is to reduce the amplitudes of doubly occupied states on the local
orbitals in the Hartree-Fock wavefunction by making use of a projection
operator $\Pi_{i} (1-gn_{i\uparrow}n_{i\downarrow})$.  Here
$n_{i\sigma}$ is the number operator for electrons on site $i$ with 
spin $\sigma$.  The variational parameter $g$ is determined by the
minimization of the ground-state energy.  Later, it has been pointed out
by Brinkman and Rice that the Gutzwiller wavefunction describes the
metal-insulator transition~\cite{br70}.  
Because the Gutzwiller method is a
nonperturbative approach, it has extensively been applied to the strongly
correlated electron systems such as the heavyfermions, high-$T_{\rm c}$
cuprates, and other transition metal oxides~\cite{geb97}.

Although the Gutzwiller ansatz approach (GA) captures the physics of electron
correlations and is useful for correlation problems, it was not so
easy to apply the method to realistic Hamiltonians.
Stollhoff and Fulde proposed an alternative method called the
local-ansatz approach (LA), which is simpler in treatment and applicable 
to realistic Hamiltonians~\cite{stoll78,stoll80}.  
The LA takes into account the excited states
created by local two-particle operators such as $\{ O_{i} \}=\{ \delta
n_{i\uparrow}\delta n_{i\downarrow} \}$, and determines their amplitudes
variationally.   Here $\delta n_{i\sigma}=
n_{i\sigma} - \langle n_{i\sigma} \rangle_{0}$, $\langle n_{i\sigma}
\rangle_{0}$ being the average electron number on site $i$ with spin
$\sigma$ in the Hartree-Fock approximation.  The LA has been applied to
many systems such as molecules, polyacetylene, transition metals,
semiconductors, and transition metal oxides~\cite{fulde95}.  

The LA is useful for understanding correlation effects
in actual materials.  The application however has been limited to the weakly 
correlated region because of the difficulty in evaluation 
of the higher-order terms
in average quantities.  In the weak
interaction limit, the Hilbert space expanded by the local operators 
is, however, not enough to describe exactly the weakly
correlated region; the LA does not reduce to the second-order 
perturbation theory in the weak correlation limit.  The same
difficulty also arises in the original Gutzwiller wavefunction even in
infinite dimensions.  In the present paper, we aim to solve these
problems in the LA introducing a new wavefunction with
momentum-dependent variational parameters, and demonstrate that the new
approach much improves the LA in the weak and intermediate correlation
regimes.  In the following, we call the new approach the MLA (the LA
with momentum dependent variational parameters).

We write down our wavefunction for the
single-band Hubbard model in \S 2.  The idea is to choose the best
local basis set obtained from the two-particle excited states 
in the momentum representation by projecting out those states
onto the local subspace and by controlling the amplitudes of the excited
states in the momentum space.  We calculate the ground-state
energy within a single-site approximation.  Using the variational
principle, we determine the momentum-dependent variational parameters.
The ground-state energy obtained from our wavefunction agrees with the
result of the second-order perturbation theory in infinite dimensions in
the weak interaction limit, and reduces to the correct atomic limit in
the case of the half-filling.

In \S 3, we present the results of numerical calculations for the
half-filled band Hubbard model in infinite dimensions to examine the
validity of the new wavefunction.  We calculate the correlation energy,
the double occupation number, the momentum distribution function, and
the quasiparticle weight as a function of the Coulomb interaction energy
parameter.  We verify that the present approach
improves both the LA and the GA in the weak and intermediate Coulomb
interaction regimes.  In particular, we demonstrate that the momentum 
distribution calculated by our wavefunction (the MLA) shows
a distinct momentum dependence, and is qualitatively 
different from those obtained by the LA and GA leading to the constant
values of the distribution function below and above the Fermi level.  
In \S 4, we present an example of applications to excitation problems.  
We have recently developed a self-consistent method~\cite{kake04-1} to
calculate the excitation spectra from the retarded Green function by
making use of the projection operator technique and the effective medium
within the coherent potential approximation
(CPA)~\cite{elliott74,shiba71}.  
The method called the projection operator method CPA (PM-CPA)~\cite{kake04-1}
is equivalent~\cite{kake02-1,kake04-2} 
to the many-body CPA~\cite{hiro77}, 
the dynamical CPA~\cite{kake92,kake02-2}, 
and the dynamical mean field 
theory~\cite{mull89,georges92,jarrell92,georges96}, 
and treats the dynamics and the static correlations
separately in the calculations.
We calculate here the excitation spectra combining our variational 
method with the PM-CPA.  
We show that the calculated spectra in the insulator regime 
quantitatively agree with the results of the numerical 
renormalization group (NRG) calculations~\cite{bulla99}.  
In the last section, we summalize our results and discuss future 
problems.
\vspace{8mm}

\section{Local Approach with Momentum Dependent Variational Parameters} 
We consider in the present paper the single-band Hubbard model defined
by 
\begin{eqnarray}
H = \sum_{i \sigma} (\epsilon_{0}-\sigma h) n_{i\sigma} 
+ \sum_{ij \sigma} t_{i j} \, a_{i \sigma}^{\dagger} a_{j \sigma} 
+ U \sum_{i} \, n_{i \uparrow} n_{i \downarrow} \ .
\label{hub}
\end{eqnarray}
Here $\epsilon_{0}$ ($h$) is the atomic level (magnetic field), 
$t_{ij}$ is the transfer integral between sites $i$ and $j$.  $U$
is the intra-atomic Coulomb energy parameter.  $ a_{i \sigma}^{\dagger}$
($ a_{i \sigma}$) denotes the creation (annihilation) operator for an
electron on site $i$ with spin $\sigma$, and 
$n_{i\sigma}=a_{i\sigma}^{\dagger} a_{i \sigma}$ is the electron density
operator on site $i$ for spin $\sigma$.

In the Hartree-Fock approximation, we replace the many-body Hamiltonian
(\ref{hub}) with an effective Hamiltonian for independent electrons,
\begin{eqnarray}
H_{0} = \sum_{ij \sigma} t_{i j\sigma} \, a_{i \sigma}^{\dagger} a_{j \sigma} 
- U \sum_{i} \, \langle n_{i \uparrow} \rangle_{0} 
\langle n_{i\downarrow} \rangle_{0} \ ,
\label{hf}
\end{eqnarray}
and approximate the ground-state wavefunction $|\Psi \rangle$ with that
of the Hartree-Fock Hamiltonian $H_{0}$, {\it i.e.}, 
$|\phi_{0} \rangle$.
Here $t_{ij\sigma}=(\epsilon_{0}+U\langle n_{i
-\sigma} \rangle_{0} - \sigma h)\delta_{ij} + t_{ij}(1-\delta_{ij})$. 
$\langle \sim \rangle_{0}$ denotes the Hartree-Fock average 
$\langle \phi_{0}| (\sim) |\phi_{0}\rangle$, and $\langle n_{i
\sigma} \rangle_{0}$ is the average electron number on site $i$ with
spin $\sigma$.
The Hamiltonian (\ref{hub}) is then expressed by a sum of the
Hartree-Fock Hamiltonian and the residual interactions as
\begin{eqnarray}
H = H_{0} + U \sum_{i} \, O_{i} \ .
\label{hub2}
\end{eqnarray}
Here $O_{i}=\delta n_{i \uparrow}\delta n_{i \downarrow} $ and 
$\delta n_{i\sigma} = n_{i\sigma} - \langle n_{i\sigma} \rangle_{0}$.

In the local-ansatz approach (LA)~\cite{stoll80}, 
we take into account the Hilbert
space created by operation of the residual interaction $\{ O_{i} \}$ onto
the Hartree-Fock state $|\phi_{0}\rangle$.  Introducing a variational
parameter $\eta^{}_{\rm \, LA}$ into the basis set $\{ O_{i} \}$,
the LA wavefunction for the ground state is written as
\begin{eqnarray}
|\Psi_{\rm LA}\rangle = \Big[ \prod_{i} (1 - \eta^{}_{\rm \, LA} O_{i})
		    \Big]|\phi_{0} \rangle \ .
\label{lawf}
\end{eqnarray}
The LA is different from the Gutzwiller ansatz wavefunction 
$|\Psi_{\rm GA}\rangle = \Big[ \prod_{i} (1 - g n_{i \uparrow} n_{i \downarrow})
		    \Big]|\phi_{0}\rangle$
in which the doubly occupied states are explicitly controlled by a
variational parameter $g$, and simplify the evaluation of the physical
quantities in the weakly correlated region.

As we have emphasized in the introduction, the LA does not lead
to the exact result in the small $U$ limit because it makes use of a
limited local subspace.  In fact, the LA wavefunction (\ref{lawf}) may
be expanded
formally in the weak interaction limit as
\begin{eqnarray}
|\Psi_{\rm LA}\rangle = |\phi_{0}\rangle + |\phi_{1}\rangle_{\rm \, LA} 
+ \cdots \ ,
\label{laexp}
\end{eqnarray}
\begin{eqnarray}
|\phi_{1}\rangle_{\rm LA} = 
- \sum_{i} \sum_{k_{1}k^{\prime}_{1}k_{2}k^{\prime}_{2}}
\langle k^{\prime}_{1}|i \rangle \langle i|k_{1} \rangle 
\langle k^{\prime}_{2}|i \rangle \langle i|k_{2} \rangle
\, \eta^{}_{\rm \, LA} \, 
\delta(a^{\dagger}_{k^{\prime}_{2}\downarrow}a_{k_{2}\downarrow})
\delta(a^{\dagger}_{k^{\prime}_{1}\uparrow}a_{k_{1}\uparrow})
|\phi_{0}\rangle \ .
\label{phi1la}
\end{eqnarray}
Here $\langle i|k \rangle = \exp (-i\boldsymbol{k} \cdot 
\boldsymbol{R}_{i}) / \sqrt{N}$ 
is an
overlap integral between the localized orbital and the Bloch state with
momentum $\boldsymbol{k}$, $\boldsymbol{R}_{i}$ denotes the atomic
position, and $N$ is the number of sites.
$ a_{k \sigma}^{\dagger}$ ($ a_{k \sigma}$) denotes the creation 
(annihilation) operator for an electron with momentum $\boldsymbol{k}$ 
and spin $\sigma$, and 
$\delta(a^{\dagger}_{k^{\prime}\sigma}a_{k\sigma})= 
a^{\dagger}_{k^{\prime}\sigma}a_{k\sigma} - \langle
a^{\dagger}_{k^{\prime}\sigma}a_{k\sigma} \rangle_{0}$.

The Rayleigh-Schr\"odinger perturbation theory, on the other hand,
yields the following form
\begin{eqnarray}
|\Psi \rangle = |\phi_{0}\rangle + |\phi_{1}\rangle + \cdots \ ,
\label{rsexp}
\end{eqnarray}
\begin{eqnarray}
|\phi_{1}\rangle = 
- \sum_{i} \sum_{k_{1}k^{\prime}_{1}k_{2}k^{\prime}_{2}}
\langle k^{\prime}_{1}|i \rangle \langle i|k_{1} \rangle 
\langle k^{\prime}_{2}|i \rangle \langle i|k_{2} \rangle
\, \eta^{(0)}_{k^{\prime}_{2}k_{2}k^{\prime}_{1}k_{1}} \, 
\delta(a^{\dagger}_{k^{\prime}_{2}\downarrow}a_{k_{2}\downarrow})
\delta(a^{\dagger}_{k^{\prime}_{1}\uparrow}a_{k_{1}\uparrow})
|\phi_{0}\rangle \ ,
\label{phi1}
\end{eqnarray}
\begin{eqnarray}
\eta^{(0)}_{k^{\prime}_{2}k_{2}k^{\prime}_{1}k_{1}} = 
-U \lim_{z \rightarrow 0}
\dfrac{f(\tilde{\epsilon}_{k_{1\uparrow}})
(1-f(\tilde{\epsilon}_{k^{\prime}_{1\uparrow}}))
f(\tilde{\epsilon}_{k_{2\downarrow}})
(1-f(\tilde{\epsilon}_{k^{\prime}_{2\downarrow}}))}
{z - \epsilon_{k^{\prime}_{1\uparrow}} + \epsilon_{k_{1\uparrow}}
- \epsilon_{k^{\prime}_{2\downarrow}} + \epsilon_{k_{2\downarrow}}
} \ .
\label{rseta}
\end{eqnarray}
Here $f(\epsilon)$ is the Fermi distribution function at zero
temperature, and $\tilde{\epsilon}_{k\sigma}=\epsilon_{k\sigma}-\mu$.
$\mu$ is the Fermi level.  $\epsilon_{k\sigma}$ is the
Hartree-Fock one-electron energy eigen value given by 
$\epsilon_{k\sigma}=\epsilon_{0}+U\langle n_{i -\sigma} \rangle_{0} +
\epsilon_{k} - \sigma h$, 
and $\epsilon_{k}$ is the Fourier transform of $t_{ij}$.

Equation (\ref{phi1}) compared with eq. (\ref{phi1la}) manifests that
one has to take into account the momentum dependence of the variational
parameters to improve the LA.
We propose in the present paper the following wavefunction with 
momentum-dependent variational parameters 
$\{ \eta_{k^{\prime}_{2}k_{2}k^{\prime}_{1}k_{1}} \}$.
\begin{eqnarray}
|\Psi\rangle = \prod_{i} (1 - \tilde{O}_{i})|\phi_{0}\rangle \ ,
\label{mla}
\end{eqnarray}
\begin{eqnarray}
\tilde{O}_{i} = \sum_{k_{1}k_{2}k^{\prime}_{1}k^{\prime}_{2}} 
\langle k^{\prime}_{1}|i \rangle \langle i|k_{1} \rangle 
\langle k^{\prime}_{2}|i \rangle \langle i|k_{2} \rangle
\eta_{k^{\prime}_{2}k_{2}k^{\prime}_{1}k_{1}} 
\delta(a^{\dagger}_{k^{\prime}_{2}\downarrow}a_{k_{2}\downarrow})
\delta(a^{\dagger}_{k^{\prime}_{1}\uparrow}a_{k_{1}\uparrow}) \ .
\label{otilde}
\end{eqnarray}
The operator $\tilde{O}_{i}$ is still localized on site $i$
because of the projection $\langle k^{\prime}_{1}|i \rangle 
\langle i|k_{1} \rangle \langle k^{\prime}_{2}|i \rangle 
\langle i|k_{2} \rangle$.  
It should be noted that  
$\tilde{O}^{\dagger}_{i} \ne \tilde{O}_{i}$ and 
$\tilde{O}_{i}\tilde{O}_{j} \ne \tilde{O}_{j}\tilde{O}_{i}$
($i \ne j$) in general.
These properties do not cause any problem when we make a single-site
approximation.  When we treat the nonlocal correlations we have to
adopt the symmetrized wavefunction in general.
The wavefunction $|\Psi\rangle$ reduces to $|\Psi_{\rm LA} \rangle$ when 
$\{ \eta_{k^{\prime}_{2}k_{2}k^{\prime}_{1}k_{1}} \}$ become
momentum-independent.

The variational parameters are determined by minimizing the ground-state
correlation energy $E_{\rm c}$. 
\begin{eqnarray}
E_{\rm c} = \langle H \rangle -  \langle H \rangle_{0} = 
\dfrac{\langle \Psi |\tilde{H}| \Psi \rangle}{\langle \Psi | \Psi \rangle} \ .
\label{corr}
\end{eqnarray}
Here $\tilde{H}=H - \langle H \rangle_{0}$.

Calculation of the correlation energy with use of the new wavefunction
is not easy in general.  But, one can evaluate it within the single-site
approximation.  As shown in Appendix A, the average $\langle \tilde{A}
\rangle$ of an operator $\tilde{A} = A -  \langle A \rangle_{0}$ with
respect to the wavefunction (\ref{mla}) is given in the single-site
approximation as
\begin{eqnarray}
\langle \tilde{A} \, \rangle = 
\sum_{i} \dfrac{\langle (1 - \tilde{O}^{\dagger}_{i}) \tilde{A}
(1 - \tilde{O}_{i}) \rangle_{0}}
{\langle (1 - \tilde{O}^{\dagger}_{i})(1 - \tilde{O}_{i}) \rangle_{0}} \ .
\label{ava}
\end{eqnarray}

By making use of the above formula, one can obtain the correlation
energy per atom. 
\begin{eqnarray}
\epsilon_{\rm c} = \dfrac{-\langle
 \tilde{O}^{\dagger}_{i}\tilde{H}\rangle_{0} -
\langle \tilde{H} \tilde{O}_{i} \rangle_{0} + 
\langle \tilde{O}^{\dagger}_{i}\tilde{H}\tilde{O}_{i}\rangle_{0}}
{1 + \langle \tilde{O}^{\dagger}_{i}\tilde{O}_{i} \rangle_{0}} \ .
\label{ec}
\end{eqnarray}
Here we assumed that all the sites are equivalent to each other and we
made use of the fact $\langle \tilde{O}^{\dagger}_{i} \rangle_{0}
= \langle \tilde{O}_{i} \rangle_{0} = 0$.

Each term in the correlation energy (\ref{ec}) can be calculated by
making use of Wick's theorem as follows.
\begin{eqnarray}
\langle \tilde{H} \tilde{O}_{i} \rangle_{0} & = &  U 
\sum_{k_{1}k_{2}k^{\prime}_{1}k^{\prime}_{2}} 
\langle k^{\prime}_{1}|i \rangle \langle i|k_{1} \rangle 
\langle k^{\prime}_{2}|i \rangle \langle i|k_{2} \rangle
\sum_{j}
\langle k_{1}|j \rangle \langle j|k^{\prime}_{1} \rangle 
\langle k_{2}|j \rangle \langle j|k^{\prime}_{2} \rangle 
\hspace{10mm}  \nonumber \\ 
& & \hspace*{15mm}  \times
\eta_{k^{\prime}_{2}k_{2}k^{\prime}_{1}k_{1}}
f(\tilde{\epsilon}_{k_{1}\uparrow})
(1-f(\tilde{\epsilon}_{k^{\prime}_{1}\uparrow}))
f(\tilde{\epsilon}_{k_{2}\downarrow})
(1-f(\tilde{\epsilon}_{k^{\prime}_{2}\downarrow})) \ ,
\label{ho}
\end{eqnarray}
\begin{eqnarray}
\langle \tilde{O}^{\dagger}_{i}\tilde{H} \rangle_{0} =
\langle \tilde{H} \tilde{O}_{i} \rangle^{\ast}_{0} \ ,
\hspace{90mm}
\label{oh}
\end{eqnarray}
\begin{eqnarray}
\langle \tilde{O}^{\dagger}_{i}\tilde{H}\tilde{O}_{i}\rangle_{0} & = & 
\sum_{k_{1}k_{2}k^{\prime}_{1}k^{\prime}_{2}} 
\langle i|k^{\prime}_{1} \rangle \langle k_{1}|i \rangle
\langle i|k^{\prime}_{2} \rangle \langle k_{2}|i \rangle \,
\eta^{\ast}_{k^{\prime}_{2}k_{2}k^{\prime}_{1}k_{1}}  \nonumber \\
& & \hspace*{-20mm} \times
f(\tilde{\epsilon}_{k_{1}\uparrow})
(1-f(\tilde{\epsilon}_{k^{\prime}_{1}\uparrow}))
f(\tilde{\epsilon}_{k_{2}\downarrow})
(1-f(\tilde{\epsilon}_{k^{\prime}_{2}\downarrow})) 
\sum_{k_{3}k_{4}k^{\prime}_{3}k^{\prime}_{4}} 
\langle k^{\prime}_{3}|i \rangle \langle i|k_{3} \rangle 
\langle k^{\prime}_{4}|i \rangle \langle i|k_{4} \rangle  \nonumber \\
& & \times
\left( \Delta E_{k^{\prime}_{2}k_{2}k^{\prime}_{1}k_{1}}
\delta_{k_{1}k_{3}}\delta_{k^{\prime}_{1}k^{\prime}_{3}}
\delta_{k_{2}k_{4}}\delta_{k^{\prime}_{2}k^{\prime}_{4}}
+ U_{k^{\prime}_{2}k_{2}k^{\prime}_{1}k_{1}
k^{\prime}_{4}k_{4}k^{\prime}_{3}k_{3}}
\right)
\eta_{k^{\prime}_{4}k_{4}k^{\prime}_{3}k_{3}} \ ,
\label{oho}
\end{eqnarray}
\begin{eqnarray}
U_{k^{\prime}_{2}k_{2}k^{\prime}_{1}k_{1}
k^{\prime}_{4}k_{4}k^{\prime}_{3}k_{3}} & = & 
U \sum_{j} [
\langle j|k_{1} \rangle \langle k_{3}|j \rangle 
f(\tilde{\epsilon}_{k_{3}\uparrow})\delta_{k^{\prime}_{1}k^{\prime}_{3}}
- \langle k^{\prime}_{1}|j \rangle \langle j|k^{\prime}_{3} \rangle
(1 - f(\tilde{\epsilon}_{k^{\prime}_{3}\uparrow})) \delta_{k_{1}k_{3}}
]  \hspace{10mm}  \nonumber \\ 
& & \hspace{5mm} \times
[\langle j|k_{2} \rangle \langle k_{4}|j \rangle 
f(\tilde{\epsilon}_{k_{4}\downarrow})\delta_{k^{\prime}_{2}k^{\prime}_{4}}
- \langle k^{\prime}_{2}|j \rangle \langle j|k^{\prime}_{4} \rangle
(1 - f(\tilde{\epsilon}_{k^{\prime}_{4}\downarrow})) \delta_{k_{2}k_{4}}
] \ ,
\label{uk}
\end{eqnarray}
\begin{eqnarray}
\langle \tilde{O}^{\dagger}_{i}\tilde{O}_{i} \rangle_{0} & = &
\dfrac{1}{N^{4}} \sum_{k_{1}k_{2}k^{\prime}_{1}k^{\prime}_{2}}
|\eta_{k^{\prime}_{2}k_{2}k^{\prime}_{1}k_{1}}|^{2}
f(\tilde{\epsilon}_{k_{1}\uparrow})
(1-f(\tilde{\epsilon}_{k^{\prime}_{1}\uparrow}))
f(\tilde{\epsilon}_{k_{2}\downarrow})
(1-f(\tilde{\epsilon}_{k^{\prime}_{2}\downarrow})) \ . 
\label{oo}
\end{eqnarray}
Here $ \Delta E_{k^{\prime}_{2}k_{2}k^{\prime}_{1}k_{1}} = 
\epsilon_{k^{\prime}_{2}\downarrow} - \epsilon_{k_{2}\downarrow}
+ \epsilon_{k^{\prime}_{1}\uparrow} - \epsilon_{k_{1}\uparrow}$
is a two-particle excitation energy.

The above expressions (\ref{ho}) and (\ref{uk}) contain nonlocal terms 
in the summation over $j$ ({\it i.e.}, $\sum_{j}$).
We thus make additional single-site approximation called the $R=0$
approximation~\cite{kajz78}.  
In eq. (\ref{ho}), for example, we have
\begin{eqnarray}
\sum_{j}
\langle k^{\prime}_{1}|i \rangle \langle i|k_{1} \rangle 
\langle k^{\prime}_{2}|i \rangle \langle i|k_{2} \rangle
\langle k_{1}|j \rangle \langle j|k^{\prime}_{1} \rangle 
\langle k_{2}|j \rangle \langle j|k^{\prime}_{2} \rangle 
= \dfrac{1}{N^{4}} \sum_{j} 
{\rm e}^{i(k_{1}+k_{2}-k^{\prime}_{1}-k^{\prime}_{2})(R_{j}-R_{i})} \ .
\label{r0}
\end{eqnarray}
The $R=0$ approximation only takes into account the local term ($j=i$)
in the above summation, so that 
$\langle \tilde{H} \tilde{O}_{i}\rangle_{0} 
(= \langle \tilde{O}^{\dagger}_{i}\tilde{H} \rangle_{0}^{\ast})$,
and 
$\langle \tilde{O}^{\dagger}_{i}\tilde{H}\tilde{O}_{i}\rangle_{0}$
reduce as follows. 
\begin{eqnarray}
\langle \tilde{H} \tilde{O}_{i} \rangle_{0} & = &  \frac{U}{N^{4}} 
\sum_{k_{1}k_{2}k^{\prime}_{1}k^{\prime}_{2}} 
f(\tilde{\epsilon}_{k_{1}\uparrow})
(1-f(\tilde{\epsilon}_{k^{\prime}_{1}\uparrow}))
f(\tilde{\epsilon}_{k_{2}\downarrow})
(1-f(\tilde{\epsilon}_{k^{\prime}_{2}\downarrow})) \,
\eta_{k^{\prime}_{2}k_{2}k^{\prime}_{1}k_{1}} \ ,
\label{hor0}
\end{eqnarray}
\begin{eqnarray}
\langle \tilde{O}^{\dagger}_{i}\tilde{H}\tilde{O}_{i}\rangle_{0} & = & 
\dfrac{1}{N^{4}} \sum_{k_{1}k_{2}k^{\prime}_{1}k^{\prime}_{2}} 
f(\tilde{\epsilon}_{k_{1}\uparrow})
(1-f(\tilde{\epsilon}_{k^{\prime}_{1}\uparrow}))
f(\tilde{\epsilon}_{k_{2}\downarrow})
(1-f(\tilde{\epsilon}_{k^{\prime}_{2}\downarrow})) \,
\eta^{\ast}_{k^{\prime}_{2}k_{2}k^{\prime}_{1}k_{1}}  
\hspace{20mm}  \nonumber \\
& &  \times
\bigg[ 
\Delta E_{k^{\prime}_{2}k_{2}k^{\prime}_{1}k_{1}} \,
\eta_{k^{\prime}_{2}k_{2}k^{\prime}_{1}k_{1}}  \nonumber \\
& &  \hspace{-10mm}
+ \dfrac{U}{N^{2}}
\Big\{
\sum_{k_{3}k_{4}}
f(\tilde{\epsilon}_{k_{3}\uparrow})f(\tilde{\epsilon}_{k_{4}\downarrow}) \,
\eta_{k^{\prime}_{2}k_{4}k^{\prime}_{1}k_{3}}
- \sum_{k_{3}k^{\prime}_{4}}
f(\tilde{\epsilon}_{k_{3}\uparrow})
(1-f(\tilde{\epsilon}_{k^{\prime}_{4}\downarrow})) \,
\eta_{k^{\prime}_{4}k_{2}k^{\prime}_{1}k_{3}}  \nonumber \\
& & \hspace*{-20mm}
- \sum_{k^{\prime}_{3}k_{4}}
(1 - f(\tilde{\epsilon}_{k^{\prime}_{3}\uparrow}))
f(\tilde{\epsilon}_{k_{4}\downarrow}) \,
\eta_{k^{\prime}_{2}k_{4}k^{\prime}_{3}k_{1}}
+ \sum_{k^{\prime}_{3}k^{\prime}_{4}}
(1 - f(\tilde{\epsilon}_{k^{\prime}_{3}\uparrow}))
(1 - f(\tilde{\epsilon}_{k_{4}\downarrow})) \,
\eta_{k^{\prime}_{4}k_{2}k^{\prime}_{3}k_{1}}
\Big\}
\bigg] \ .
\label{ohor0}
\end{eqnarray}

Variational parameters 
$\{ \eta_{k^{\prime}_{2}k_{2}k^{\prime}_{1}k_{1}} \}$
are obtained by minimizing the correlation energy $\epsilon_{\rm c}$, 
{\it i.e.}, eq. (\ref{ec}) with eqs. (\ref{oo}), (\ref{hor0}), and
(\ref{ohor0}).
The self-consistent equations for 
$\{ \eta_{k^{\prime}_{2}k_{2}k^{\prime}_{1}k_{1}} \}$
in the single-site approximation are given as follows.
\begin{eqnarray}
(\Delta E_{{k^{\prime}_{2}k_{2}k^{\prime}_{1}k_{1}}} - \epsilon_{\rm c})
\eta_{k^{\prime}_{2}k_{2}k^{\prime}_{1}k_{1}}  \hspace{5mm} \nonumber \\
& &  \hspace{-50mm}
+ \dfrac{U}{N^{2}}
\Big[
\sum_{k_{3}k_{4}}
f(\tilde{\epsilon}_{k_{3}\uparrow})f(\tilde{\epsilon}_{k_{4}\downarrow})
\eta_{k^{\prime}_{2}k_{4}k^{\prime}_{1}k_{3}}
- \sum_{k_{3}k^{\prime}_{4}}
f(\tilde{\epsilon}_{k_{3}\uparrow})
(1-f(\tilde{\epsilon}_{k^{\prime}_{4}\downarrow}))
\eta_{k^{\prime}_{4}k_{2}k^{\prime}_{1}k_{3}}  \nonumber \\
& & \hspace*{-50mm}
- \sum_{k^{\prime}_{3}k_{4}}
(1 - f(\tilde{\epsilon}_{k^{\prime}_{3}\uparrow}))
f(\tilde{\epsilon}_{k_{4}\downarrow})
\eta_{k^{\prime}_{2}k_{4}k^{\prime}_{3}k_{1}}
+ \sum_{k^{\prime}_{3}k^{\prime}_{4}}
(1 - f(\tilde{\epsilon}_{k^{\prime}_{3}\uparrow}))
(1 - f(\tilde{\epsilon}_{k_{4}\downarrow}))
\eta_{k^{\prime}_{4}k_{2}k^{\prime}_{3}k_{1}}
\Big] = U \, . \ \ \
\label{eqeta}
\end{eqnarray}
Note that $\eta_{k^{\prime}_{2}k_{2}k^{\prime}_{1}k_{1}}$ should vanish
when $U \rightarrow 0$.  Thus in the weak $U$ limit, one can omit
the second term at the l.h.s. (left-hand-side) of eq. (\ref{eqeta}).
We then obtain the solution in the weak $U$ limit as
\begin{eqnarray}
\eta_{k^{\prime}_{2}k_{2}k^{\prime}_{1}k_{1}} = 
\dfrac{U}{\Delta E_{{k^{\prime}_{2}k_{2}k^{\prime}_{1}k_{1}}}} \ .
\label{etaweak}
\end{eqnarray}
In the atomic limit, 
$\Delta E_{{k^{\prime}_{2}k_{2}k^{\prime}_{1}k_{1}}}=0$.
We find then a $k$-independent solution being identical with the LA.
\begin{eqnarray}
\eta^{}_{\rm \, LA} = 
\frac{\displaystyle -\langle O_{i}\tilde{H}O_{i}\rangle_{0} + 
\sqrt{\langle O_{i}\tilde{H}O_{i}\rangle_{0}^{2}+
4\langle O_{i}\tilde{H}\rangle^{2}_{0}\langle O_{i}^{2}\rangle_{0}}
}
{2\langle O_{i}\tilde{H}\rangle_{0}\langle O_{i}^{2}\rangle_{0}
} \ .
\label{etala}
\end{eqnarray}

It is not easy to find the solution of eq. (\ref{eqeta}) for the
intermediate strength of Coulomb interaction $U$.  We therefore consider
an approximate solution which interpolates between the weak and the
atomic limits.  Note that the second term at the l.h.s. of
eq. (\ref{eqeta}) do not affect the solution in the weakly correlated
limit as we have mentioned.
Therefore we approximate 
$\{ \eta_{k^{\prime}_{2}k_{2}k^{\prime}_{1}k_{1}} \}$ in the second term
with the momentum-independent parameter $\eta$ which is suitable for the
atomic region.  We have then an approximate solution as follows.
\begin{eqnarray}
\eta_{k^{\prime}_{2}k_{2}k^{\prime}_{1}k_{1}} = 
\dfrac{U[1 - \eta (1 - 2 \langle n_{i\uparrow} \rangle_{0}) 
(1 - 2 \langle n_{i\downarrow} \rangle_{0})]}
{\Delta E_{{k^{\prime}_{2}k_{2}k^{\prime}_{1}k_{1}}} - \epsilon_{\rm c}} \ .
\label{etaint}
\end{eqnarray}
The best value of $\eta$ should be determined variationally, but we make
use of that in the LA for simplicity.  Furthermore, we approximate
$\epsilon_{\rm c}$ in the denominator of eq. (\ref{etaint}) with the
correlation energy in the LA.

Substituting the variational parameters (\ref{etaint}) into
eq. (\ref{ec}), we obtain the ground-state correlation energy.  The
each element in the energy is given as follows.
\begin{eqnarray}
\langle \tilde{H} \tilde{O}_{i}\rangle_{0} 
& =  & \langle \tilde{O}^{\dagger}_{i}\tilde{H} \rangle_{0}^{\ast}
\nonumber \\
& = & U^{2} [1 - \eta (1 - 2 \langle n_{i\uparrow} \rangle_{0}) 
(1 - 2 \langle n_{i\downarrow} \rangle_{0})]  \nonumber \\
& & \hspace{-10mm} \times
\int \dfrac{ \big[\prod_{n} d\epsilon_{n}\big]
\rho_{\uparrow}(\epsilon_{1})
\rho_{\uparrow}(\epsilon_{2})\rho_{\downarrow}(\epsilon_{3})
\rho_{\downarrow}(\epsilon_{4})
f(\epsilon_{1})(1-f(\epsilon_{2}))f(\epsilon_{3})(1-f(\epsilon_{4}))
}
{ 
\epsilon_{4} - \epsilon_{3} + \epsilon_{2} - \epsilon_{1} -
\epsilon_{\rm c}
} \ ,
\label{ho2}
\end{eqnarray}
\begin{eqnarray}
\langle \tilde{O}^{\dagger}_{i} \tilde{H} \tilde{O}_{i} \rangle_{0} = 
\langle \tilde{O}^{\dagger}_{i} \tilde{H}_{0} \tilde{O}_{i} \rangle_{0} 
+ U \langle \tilde{O}^{\dagger}_{i} O_{i} \tilde{O}_{i} \rangle_{0} \ ,
\hspace{67mm}
\label{oho2}
\end{eqnarray}
\begin{eqnarray}
\langle \tilde{O}^{\dagger}_{i} \tilde{H}_{0} \tilde{O}_{i} \rangle_{0} 
& = & U^{2} [1 - \eta (1 - 2 \langle n_{i\uparrow} \rangle_{0}) 
(1 - 2 \langle n_{i\downarrow} \rangle_{0})]^{2}   \nonumber \\
& & \hspace{0mm} \times
\int \Big[\prod_{n} d\epsilon_{n}\Big]
\rho_{\uparrow}(\epsilon_{1})
\rho_{\uparrow}(\epsilon_{2})\rho_{\downarrow}(\epsilon_{3})
\rho_{\downarrow}(\epsilon_{4})  \nonumber \\
& &  \times
f(\epsilon_{1})(1-f(\epsilon_{2}))f(\epsilon_{3})(1-f(\epsilon_{4}))
\dfrac{\epsilon_{4} - \epsilon_{3} + \epsilon_{2} - \epsilon_{1}}
{(\epsilon_{4} - \epsilon_{3} + \epsilon_{2} - \epsilon_{1} -
\epsilon_{\rm c})^{2}
} \ ,   \hspace{15mm}
\label{oh0o2}
\end{eqnarray}
\begin{eqnarray}
\langle \tilde{O}^{\dagger}_{i} O_{i} \tilde{O}_{i} \rangle_{0} & = &
U^{2} [1 - \eta (1 - 2 \langle n_{i\uparrow} \rangle_{0}) 
(1 - 2 \langle n_{i\downarrow} \rangle_{0})]^{2}  
\hspace{15mm} \nonumber \\
& & \hspace{-10mm} \times
\int \Big[\prod_{n} d\epsilon_{n}\Big]
\rho_{\uparrow}(\epsilon_{1})
\rho_{\uparrow}(\epsilon_{2})\rho_{\downarrow}(\epsilon_{3})
\rho_{\downarrow}(\epsilon_{4})
\dfrac{f(\epsilon_{1})(1-f(\epsilon_{2}))f(\epsilon_{3})(1-f(\epsilon_{4}))}
{\epsilon_{4} - \epsilon_{3} + \epsilon_{2} - \epsilon_{1} -
\epsilon_{\rm c}}  \hspace{10mm} \nonumber \\
& & \hspace{-20mm} \times \bigg[
\int \dfrac{d\epsilon_{5}d\epsilon_{6}
\rho_{\uparrow}(\epsilon_{5})\rho_{\downarrow}(\epsilon_{6})
f(\epsilon_{5})f(\epsilon_{6})}
{\epsilon_{4} - \epsilon_{6} + \epsilon_{2} - \epsilon_{5} -
\epsilon_{\rm c}}
- \int \dfrac{d\epsilon_{5}d\epsilon_{6}
\rho_{\uparrow}(\epsilon_{5})\rho_{\downarrow}(\epsilon_{6})
f(\epsilon_{5})(1-f(\epsilon_{6}))}
{\epsilon_{6} - \epsilon_{3} + \epsilon_{2} - \epsilon_{5} -
\epsilon_{\rm c}}  \nonumber \\
& & \hspace{-25mm} 
- \int \dfrac{d\epsilon_{5}d\epsilon_{6}
\rho_{\uparrow}(\epsilon_{5})\rho_{\downarrow}(\epsilon_{6})
(1-f(\epsilon_{5}))f(\epsilon_{6})}
{\epsilon_{4} - \epsilon_{6} + \epsilon_{5} - \epsilon_{1} -
\epsilon_{\rm c}}
+ \!\! \int \dfrac{d\epsilon_{5}d\epsilon_{6}
\rho_{\uparrow}(\epsilon_{5})\rho_{\downarrow}(\epsilon_{6})
(1-f(\epsilon_{5}))(1-f(\epsilon_{6}))}
{\epsilon_{6} - \epsilon_{3} + \epsilon_{5} - \epsilon_{1} -
\epsilon_{\rm c}}
\bigg] ,  \hspace{7mm}
\label{ooo02}
\end{eqnarray}
\begin{eqnarray}
\langle \tilde{O}^{\dagger}_{i}\tilde{O}_{i} \rangle_{0} & = &
U^{2} [1 - \eta (1 - 2 \langle n_{i\uparrow} \rangle_{0}) 
(1 - 2 \langle n_{i\downarrow} \rangle_{0})]^{2}  
\hspace{15mm} \nonumber \\
& & \hspace{-10mm} \times
\int \Big[\prod_{n} d\epsilon_{n}\Big]
\rho_{\uparrow}(\epsilon_{1})
\rho_{\uparrow}(\epsilon_{2})\rho_{\downarrow}(\epsilon_{3})
\rho_{\downarrow}(\epsilon_{4})
\dfrac{f(\epsilon_{1})(1-f(\epsilon_{2}))f(\epsilon_{3})(1-f(\epsilon_{4}))}
{(\epsilon_{4} - \epsilon_{3} + \epsilon_{2} - \epsilon_{1} -
\epsilon_{\rm c})^{2}} \ . \hspace{20mm}
\label{oo2}
\end{eqnarray}
Here $\rho_{\sigma}(\epsilon)$ is the density of states for the
Hartree-Fock one-electron energy eigen values measured from the Fermi 
level.

Electron number 
$\langle n_{i} \rangle (= \sum_{\sigma} \langle n_{i\sigma} \rangle)$,
the momentum distribution $\langle n_{k\sigma} \rangle$, and the 
double occupation number $\langle n_{i\uparrow}n_{i\downarrow} \rangle$
are obtained from $\partial \langle H \rangle / \partial \epsilon$, 
$\partial \langle H \rangle / \partial \hat{\epsilon}_{k\sigma}$, and
$\partial \langle H \rangle / \partial U_{i}$, 
respectively.  Here $\hat{\epsilon}_{k\sigma} = \epsilon_{k} - \sigma h$
and $\epsilon_{k}$ is the Fourier transform of $t_{ij}$.
Making use of the single-site energy (\ref{ec}) and the Feynman-Hellmann
theorem, we obtain the following expressions.
\begin{eqnarray}
\langle n_{i} \rangle = \langle n_{i} \rangle_{0} + 
\dfrac{\langle \tilde{O}_{i} \tilde{n}_{i} \tilde{O}_{i} \rangle_{0}}
{1+\langle \tilde{O}^{\dagger}_{i} \tilde{O}_{i} \rangle_{0}} \ ,
\label{ni}
\end{eqnarray}
\begin{eqnarray}
\langle n_{k\sigma} \rangle = \langle n_{k\sigma} \rangle_{0} + 
\dfrac{N \langle \tilde{O}_{i} \tilde{n}_{k\sigma} \tilde{O}_{i} \rangle_{0}}
{1+\langle \tilde{O}^{\dagger}_{i} \tilde{O}_{i} \rangle_{0}} \ ,
\label{nk}
\end{eqnarray}
\begin{eqnarray}
\langle n_{i\uparrow}n_{i\downarrow} \rangle = 
\langle n_{i\uparrow} \rangle_{0} \langle n_{i\downarrow} \rangle_{0}
+ \dfrac{-\langle \tilde{O}^{\dagger}_{i} O_{i} \rangle_{0} 
- \langle O_{i} \tilde{O}_{i} \rangle_{0} 
+ \langle \tilde{O}^{\dagger}_{i} O_{i} \tilde{O}_{i} \rangle_{0}
+ \sum_{\sigma} \langle n_{i-\sigma} \rangle_{0} 
\langle \tilde{O}^{\dagger}_{i} \tilde{n}_{i\sigma} \tilde{O}_{i} \rangle_{0}
}
{1+\langle \tilde{O}^{\dagger}_{i} \tilde{O}_{i} \rangle_{0}} \ .
\label{dble}
\end{eqnarray}
Here $\tilde{n}_{i} = n_{i} -  \langle n_{i} \rangle_{0}$,
$\tilde{n}_{k\sigma} = n_{k\sigma} -  \langle n_{k\sigma} \rangle_{0}$, and
\begin{eqnarray}
\langle \tilde{O}^{\dagger}_{i} \tilde{n}_{i\sigma} \tilde{O}_{i} 
\rangle_{0} & = & 
2 U^{2} [1 - \eta (1 - 2 \langle n_{i\uparrow} \rangle_{0}) 
(1 - 2 \langle n_{i\downarrow} \rangle_{0})]^{2}  
\hspace{15mm} \nonumber \\
& & \hspace{-10mm} \times
\int \Big[\prod_{n} d\epsilon_{n}\Big]
\dfrac{
\rho_{-\sigma}(\epsilon_{1})
\rho_{-\sigma}(\epsilon_{2})\rho_{\sigma}(\epsilon_{3})
\rho_{\sigma}(\epsilon_{4})\rho_{\sigma}(\epsilon_{5})
f(\epsilon_{1})(1-f(\epsilon_{2}))f(\epsilon_{3})(1-f(\epsilon_{4}))}
{\epsilon_{4} - \epsilon_{3} + \epsilon_{2} - \epsilon_{1} -
\epsilon_{\rm c}}
\hspace{5mm} \nonumber \\
& & \hspace{20mm} \times \bigg[
\dfrac{1-f(\epsilon_{5})}
{\epsilon_{5} - \epsilon_{3} + \epsilon_{2} - \epsilon_{1} -
\epsilon_{\rm c}}
- \dfrac{f(\epsilon_{5})}
{\epsilon_{4} - \epsilon_{5} + \epsilon_{2} - \epsilon_{1} -
\epsilon_{\rm c}}
\bigg] \ ,
\label{ono2}
\end{eqnarray}
\begin{eqnarray}
N \langle \tilde{O}^{\dagger}_{i} \tilde{n}_{k\sigma} \tilde{O}_{i} 
\rangle_{0} & = &
U^{2} [1 - \eta (1 - 2 \langle n_{i\uparrow} \rangle_{0}) 
(1 - 2 \langle n_{i\downarrow} \rangle_{0})]^{2}  \nonumber \\
& & \hspace{-5mm} \times
\bigg[
(1 - f(\epsilon_{k\sigma})) \int 
\dfrac{ d\epsilon_{1}d\epsilon_{2}d\epsilon_{3}
\rho_{-\sigma}(\epsilon_{1})
\rho_{-\sigma}(\epsilon_{2})\rho_{\sigma}(\epsilon_{3})
f(\epsilon_{1})(1-f(\epsilon_{2}))f(\epsilon_{3})}
{(\epsilon_{1} - \epsilon_{2} + \epsilon_{k\sigma} - \epsilon_{3} -
\epsilon_{\rm c})^{2}}  \nonumber \\
& & \hspace{-15mm} 
- f(\epsilon_{k\sigma}) \int
\dfrac{ d\epsilon_{1}d\epsilon_{2}d\epsilon_{3}
\rho_{-\sigma}(\epsilon_{1})
\rho_{-\sigma}(\epsilon_{2})\rho_{\sigma}(\epsilon_{3})
f(\epsilon_{1})(1-f(\epsilon_{2}))(1 - f(\epsilon_{3}))}
{(\epsilon_{1} - \epsilon_{2} + \epsilon_{3} - \epsilon_{k\sigma} - 
\epsilon_{\rm c})^{2}}
\bigg] \ , \hspace{10mm}
\label{onko2}
\end{eqnarray}
\begin{eqnarray}
\langle \tilde{O}^{\dagger}_{i} O_{i} \rangle_{0} 
+ \langle O_{i} \tilde{O}_{i} \rangle_{0} & = &
2 U [1 - \eta (1 - 2 \langle n_{i\uparrow} \rangle_{0}) 
(1 - 2 \langle n_{i\downarrow} \rangle_{0})]  \nonumber \\
& & \hspace{-20mm} \times
\int \dfrac{ \big[\prod_{n} d\epsilon_{n}\big]
\rho_{\uparrow}(\epsilon_{1})
\rho_{\uparrow}(\epsilon_{2})\rho_{\downarrow}(\epsilon_{3})
\rho_{\downarrow}(\epsilon_{4})
f(\epsilon_{1})(1-f(\epsilon_{2}))f(\epsilon_{3})(1-f(\epsilon_{4}))}
{\epsilon_{4} - \epsilon_{3} + \epsilon_{2} - \epsilon_{1} -
\epsilon_{\rm c}} \ . \hspace{10mm}
\label{otildeo2}
\end{eqnarray}

It should be noted that the expressions of these physical quantities
consist of the multiple integrals up to the 6-folds.  One can reduce
these integrals up to the 2-folds using the Laplace transform.  Their
expressions are given in Appendix B.
%
%
\begin{figure}
\includegraphics{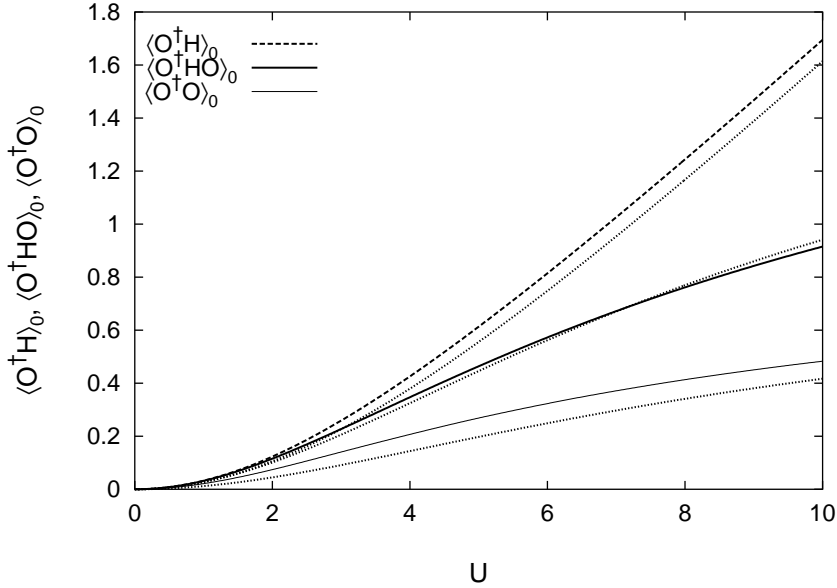}%
\caption{\label{figoho}
Calculated elements 
$\langle \tilde{O}^{\dagger}_{i} \tilde{H} \rangle_{0}$ (dashed curve), 
$\langle \tilde{O}^{\dagger}_{i} \tilde{H} \tilde{O}_{i} \rangle_{0}$
(solid curve),
and $\langle \tilde{O}^{\dagger}_{i} \tilde{O}_{i} \rangle_{0}$ 
(thin solid curve) as a function of the Coulomb interaction parameter $U$.
Corresponding curves in the LA are shown by dotted curves.
}
\end{figure}
%
%

\section{Numerical Example: Half-Filled Hubbard Model}
We have performed the numerical calculations of the half-filled band
Hubbard model in order to examine the properties of the local ansatz
with momentum-dependent variational parameters (MLA).
We consider here the hypercubic lattice in infinite dimensions, where
the single-site approximation works best.
The density of states for noninteracting system on the hypercubic
lattice is given by 
$\rho(\epsilon) = (1/\sqrt{\pi}) \exp (-\epsilon^{2})$, 
in which the energy unit is chosen to be $\int d\epsilon \rho (\epsilon)
\epsilon^{2} = 1/2$.

Figure  \ref{figoho} shows the curves for 
$\langle \tilde{O}^{\dagger}_{i} \tilde{H} \rangle_{0}$, 
$\langle \tilde{O}^{\dagger}_{i} \tilde{H} \tilde{O}_{i} \rangle_{0}$,
and $\langle \tilde{O}^{\dagger}_{i} \tilde{O}_{i} \rangle_{0}$ 
in the correlation energy as a function of the Coulomb interaction
energy parameter $U$.  These are proportional to $U^{2}$ in the small
$U$ limit.
The energy $\langle \tilde{O}^{\dagger}_{i} \tilde{H} \rangle_{0}$
agrees with 
$\langle \tilde{O}^{\dagger}_{i} \tilde{H} \tilde{O}_{i} \rangle_{0}$
in the small $U$ limit as is verified from eqs. (\ref{ho2}) and 
(\ref{oho2}).
This implies that $\epsilon_{\rm c} \approx - 
\langle \tilde{O}^{\dagger}_{i} \tilde{H} \rangle_{0}$
in the small $U$ limit.
For large $U$, the energy 
$\langle \tilde{O}^{\dagger}_{i} \tilde{H} \tilde{O}_{i} \rangle_{0}$
becomes smaller than 
$\langle \tilde{O}^{\dagger}_{i} \tilde{H} \rangle_{0}$
because the former changes with 
$\{ \eta_{k^{\prime}_{2}k_{2}k^{\prime}_{1}k_{1}} \}$
in a quadratic way, while the latter linearly depends on $U$ as 
$U \eta_{k^{\prime}_{2}k_{2}k^{\prime}_{1}k_{1}}$, and
because $\{ \eta_{k^{\prime}_{2}k_{2}k^{\prime}_{1}k_{1}} \}$
saturate when $U$ is large.  The renormalization contribution 
$\langle \tilde{O}^{\dagger}_{i} \tilde{O}_{i} \rangle_{0}$
tends to saturate with increasing $U$ because of the same reason.
As a consequence, the energy
$-\langle \tilde{O}^{\dagger}_{i} \tilde{H} \rangle_{0}$
forms the leading term in the correlation energy
even for large $U$.

The same quantities in the LA are obtained by the replacements 
$\eta_{k^{\prime}_{2}k_{2}k^{\prime}_{1}k_{1}} \rightarrow 
\eta^{}_{\rm \, LA}$;
$\langle \tilde{O}^{\dagger}_{i} \tilde{H} \rangle_{0} \rightarrow 
\eta^{}_{\rm \, LA} \langle O_{i} \tilde{H} \rangle_{0}$,
$\langle \tilde{O}^{\dagger}_{i} \tilde{H} \tilde{O}_{i} \rangle_{0} 
\rightarrow \eta^{2}_{\rm \, LA} \langle O_{i} \tilde{H} O_{i}
\rangle_{0}$, and
$\langle \tilde{O}^{\dagger}_{i} \tilde{O}_{i} \rangle_{0} 
\rightarrow \eta^{2}_{\rm \, LA} \langle O^{2}_{i} \rangle_{0}$ in eqs.
(\ref{ho2}), (\ref{oho2}), and (\ref{oo2}).  Their expressions are
calculated analytically in the present case as 
$\eta^{}_{\rm \, LA} \langle O_{i} \tilde{H} \rangle_{0} = 
\eta^{}_{\rm \, LA} U/16$, 
$\eta^{2}_{\rm \, LA} \langle O_{i} \tilde{H} O_{i} \rangle_{0} = 
\eta^{2}_{\rm \, LA}/4\sqrt{\pi}$, 
$\eta^{2}_{\rm \, LA} \langle O^{2}_{i} \rangle_{0} = 
\eta^{2}_{\rm \, LA}/16$, 
and  
\begin{eqnarray}
\eta^{}_{\rm \, LA} = \dfrac{-\dfrac{1}{\sqrt{\pi}} + \sqrt{\dfrac{1}{\pi} 
+ \dfrac{U^{2}}{16}}} {\dfrac{U}{32}} \ .
\label{etala2}
\end{eqnarray}
These quantities in the LA are also presented in Fig. \ref{figoho} by
dotted curves.  The results in the LA describe those in the MLA rather
well over a wide range of the Coulomb interaction.
It should be noted that the curves in the LA deviate from those in the MLA
even in the small $U$ limit.  A remarkable point is that
$\langle \tilde{O}^{\dagger}_{i} \tilde{H} \rangle_{0}$ is larger than
that in the LA.  This lowers the ground-state energy of the MLA.
%
%
\begin{figure}
\includegraphics{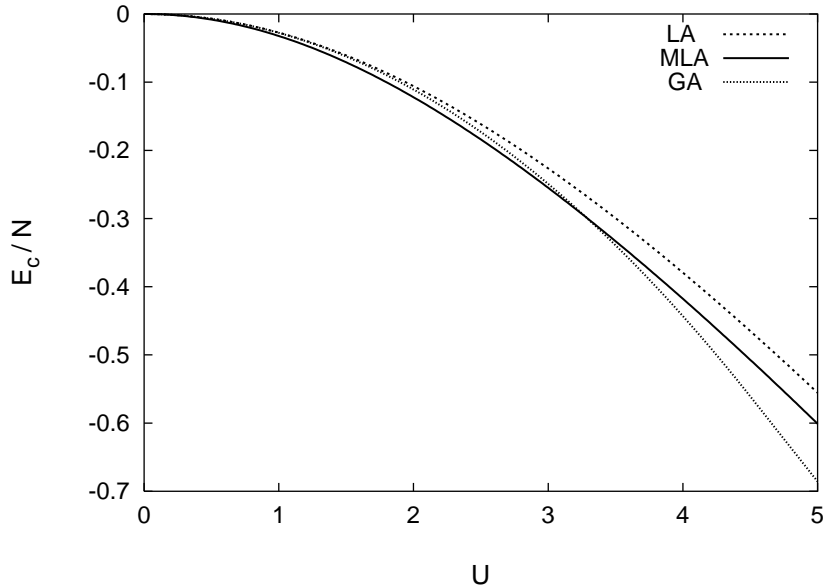}%
\caption{\label{figec}
The correlation energies vs. Coulomb interaction energy parameter $U$ 
in the LA (dashed curve), the MLA (solid curve), and the GA (dotted curve).
}
\end{figure}
%
%

Figure \ref{figec} shows calculated correlation energy as a function of
Coulomb interaction.  The energy in the MLA is lower than that of the LA
over all Coulomb interaction energy parameters $U$, verifying an
improvement of the wavefunction.
Moreover, the MLA wavefunction leads to the correlation energy lower
than that
of the original Gutzwiller Ansatz (GA) for the Coulomb interaction
energy parameter $U \le U^{\ast}=3.28$.  We can expect that the MLA
yields better results in the interaction range $[0, U^{\ast}]$, while
the original GA should be better for $U \ge U^{\ast}$. 

\begin{figure}
\includegraphics{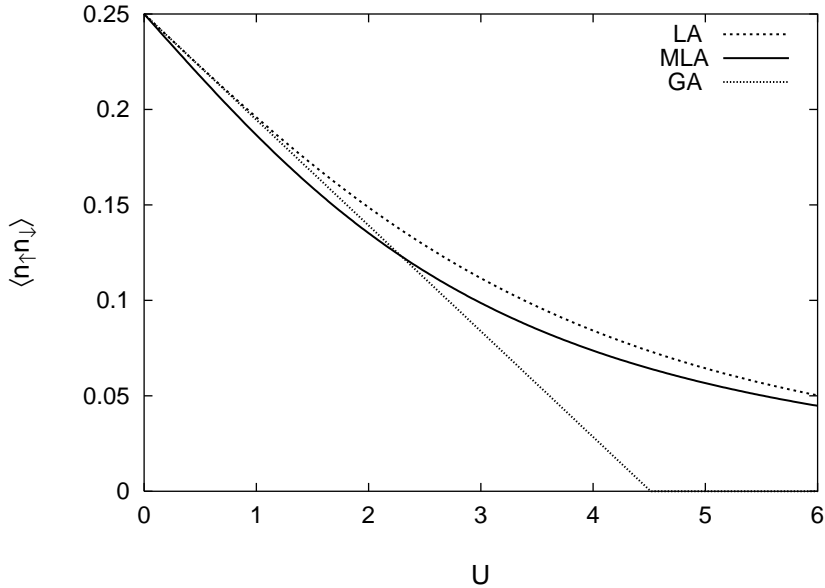}%
\caption{\label{figdbl}
The double occupation number vs. Coulomb interaction energy $U$ curves 
in the LA (dashed curve), the MLA (solid curve), and the GA (dotted curve).
}
\end{figure}

We present the double occupation number vs. Coulomb interaction curves in
Fig. \ref{figdbl}.  The double occupancy in the uncorrelated limit is
$1/4$, and decreases with increasing Coulomb interaction $U$.  Both the
LA and the MLA yield $\langle n_{i\uparrow}n_{i\downarrow} \rangle = 0$
in the limit $U=\infty$.  The MLA suppresses 
$\langle n_{i\uparrow}n_{i\downarrow} \rangle$ of the LA
typically by about 10 \% in the
intermediate regime of Coulomb interaction.
The double occupation number in the GA linearly decreases with 
increasing $U$ and causes the metal-insulator transition at 
$U_{\rm c2}=8/\sqrt{\pi}(=4.51)$.  The GA underestimates the double
occupancy in the insulator regime because 
$\langle n_{i\uparrow}n_{i\downarrow} \rangle$
should be finite even beyond $U_{\rm c2}$ due to virtual exchange of
electrons between the nearest neighbor atoms.  Present result of the MLA
indicates that the GA overestimates 
$\langle n_{i\uparrow}n_{i\downarrow} \rangle$
for small $U (\lesssim 2)$ and underestimates it at $U \sim 3$.
%
%
%
%

\begin{figure}
\includegraphics{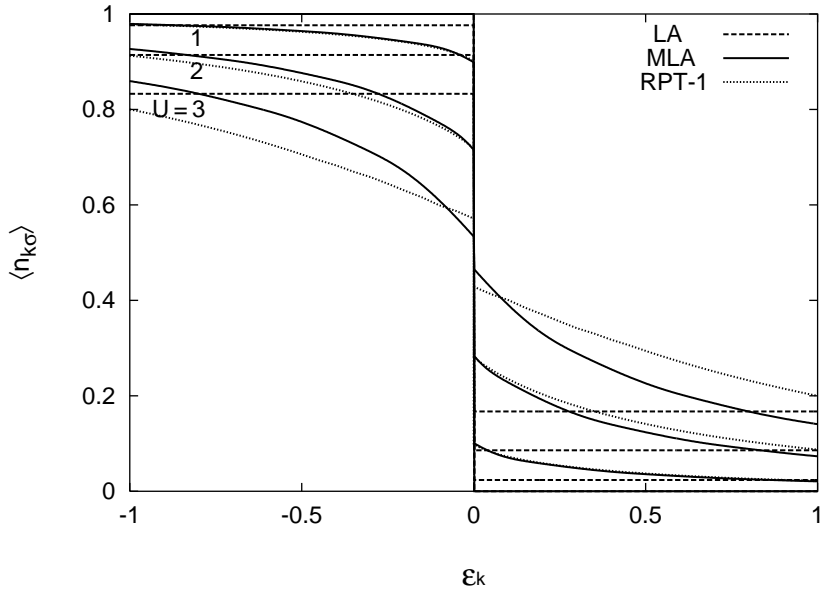}%
\caption{\label{fignk}
The momentum distribution curves as a function of energy $\epsilon_{k}$ 
for various Coulomb interaction energy parameters $U$. 
The results of the LA: dashed curves, the MLA: solid curves, and 
the RPT-1: dotted curves~\cite{kake04-1}.
}
\end{figure}

The difference between the LA and the MLA is seen more clearly in the
momentum-dependent quantities.  Figure \ref{fignk} shows the momentum
distribution in various approximations.  The distributions in the LA are
constant below and above the Fermi level irrespective of $U$.  This
behavior is also found in the GA~\cite{gutz64}.  
The MLA shows a distinct momentum dependence 
of $\langle n_{k\sigma} \rangle$ via the energy $\epsilon_{k}$. 
The results are in good 
agreement with those in the previous results of the RPT-1 (The
first-order approximation in the renormalized perturbation theory) in the
projection operator method CPA~\cite{kake04-1}.  
The latter is exact up to the second
order in $U$, and reproduce the Hubbard III approximation in the large
$U$ region.
%
%
%
%

\begin{figure}
\includegraphics{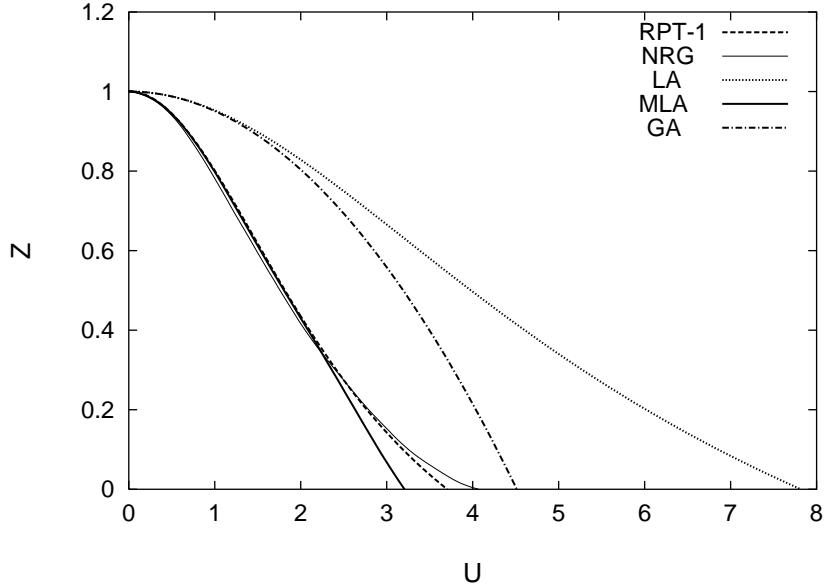}%
\caption{\label{figzu}
Quasiparticle-weight vs. Coulomb interaction curves in various theories.
The RPT-1: dashed curve, the NRG: thin solid curve~\cite{bulla99}, 
the LA: dotted curve, the MLA: solid curve, and the GA: dot-dashed
 curve~\cite{br70}. 
}
\end{figure}

The jump at the Fermi level in the momentum distribution gives us the
quasiparticle weight $Z$ ({\it i.e.} the inverse effective mass).
Calculated quasiparticle weight vs. Coulomb interaction curves are shown
in Fig. \ref{figzu}.  The quasiparticle weight in the LA changes as
$Z=(1-3\eta^{2}_{\rm \, LA}/16)/(1+\eta^{2}_{\rm \, LA}/16)$ and
vanishes at $U_{\rm c2}(\rm LA) = 24/\sqrt{3\pi} \, (=7.82)$.  
In the GA~\cite{br70}, the quasiparticle weight changes as
$Z=1-(U/U_{\rm c2})^{2}$.  The curve in the GA agrees with the LA curve for
small $U$.  But it deviates from the LA when $U$ becomes large, 
and vanishes at
$U_{\rm c2}(\rm GA)=8/\sqrt{\pi} \, (=4.51)$.  It should be noted that the GA
curve strongly deviates from the curve in the NRG~\cite{bulla99} 
which is considered to
be the best at present.  The quasiparticle weight in the MLA much
improves the LA; it is close to the curve in the NRG up to $U \approx
2.5$, and vanishes at $U_{\rm c2}(\rm MLA)=3.21$.  
The latter should be compared
with $U_{\rm c2}(\rm NRG)=4.10$.  We note that the wavefunction itself
does not show the metal-insulator transition at $U_{\rm c2}$ in the present 
approximation because the approximate
expression of variational parameters (\ref{etaint})
has no singularity at finite value of $U$.
In this sense, the calculated
$Z$ and wavefunction are not self-consistent in the present
approximation. 
The values of $Z$ obtained by the LA and MLA should be regarded
as an estimate from the metallic side.  
%
%
%
%

\section{Application to Excitation Spectra}
We demonstrate in this section that the present approach is also useful for
understanding the correlation effects on excitation spectra.  This
can be made by combining the MLA with the projection operator method
(see, for example, Chap. 6 in Ref. 7).
The projection technique treats the dynamics of electrons and the 
static average separately, and the latter can be calculated by using the
wavefunction method.

We adopt again the half-filled band Hubbard model in infinite
dimensions, and apply the projection operator CPA method 
(PM-CPA)~\cite{kake04-1}.
In this methods, we describe the single-particle excitations by means of
the Fourier transform of the retarded Green
function, $(a^{\dagger}_{i\sigma}|(z-L)^{-1}a^{\dagger}_{j\sigma})$.
Here the Liouville operator $L$ defined by $LA = [H,A]_{-}$
for an operator $A$ describes the dynamics of electrons,  
$z=\omega+i\delta$, $\delta$ is an infinitesimal positive number, and
the inner product between the operators $A$ and $B$ is defined by
$(A|B)=\langle [A^{\dagger},B]_{+} \rangle$.
We approximate in the PM-CPA the operator $L$ 
by an energy dependent Liouvillean $\tilde{L}(z)$
for an effective Hamiltonian with a coherent potential 
$\tilde{\Sigma}(z)$, 
$\sum_{i\sigma}\tilde{\Sigma}(z) n_{i\sigma} +
\sum_{ij \sigma} t_{i j} \, a_{i \sigma}^{\dagger} a_{j \sigma}$.  
The Green function for $\tilde{L}(z)$ is given by 
\begin{eqnarray}
F(z) = 
\int \frac{\rho(\epsilon) \, d\epsilon}
{z - \tilde{\Sigma}(z)-\epsilon} \ .
\label{cohg2}
\end{eqnarray}
Here $\rho(\epsilon)$ is the density of states (DOS) per site for the
noninteracting system.

To obtain the coherent potential $\tilde{\Sigma}(z)$, we consider an
impurity system with Coulomb interaction on a site embedded in the
coherent potential.  The site-diagonal impurity Green function for the
system is then given by 
\begin{eqnarray}
G^{(i)}(z) = 
\left( F(z)^{-1} - \tilde{\Lambda}^{(i)}(z) + \tilde{\Sigma}(z)
\right)^{-1} \ .
\label{gimp2}
\end{eqnarray}
The self-energy $\tilde{\Lambda}^{(i)}(z)$ is expressed as follows
for the half-filled band according to the RPT (Renormalized Perturbation
Theory)~\cite{kake04-1}.
\begin{eqnarray}
\tilde{\Lambda}^{(i)}(z) =  
\frac{\displaystyle U^{2} \overline{G}^{(i)}_{0}(z)}
{\displaystyle 1 + 4\tilde{\Sigma}(z) \, 
\overline{G}^{(i)}_{0}(z)} \ ,
\label{rself0}
\end{eqnarray}
\begin{eqnarray}
\overline{G}^{(i)}_{0}(z)= 
(A^{\dagger}_{i\sigma}|(z-\overline{L}_{0}(z)-
\overline{L}^{(i)}_{I}(z)\overline{Q})^{-1} 
A^{\dagger}_{i\sigma}) \ .
\label{g00-2}
\end{eqnarray}
Here $A^{\dagger}_{i\sigma} \equiv a^{\dagger}_{i\sigma}\delta n_{i-\sigma}$
is an atomic operator expanded by the
intra-atomic Coulomb interaction.
$\overline{L}_{0}(z)=Q\tilde{L}(z)Q$ and
$\overline{L}^{(i)}_{I}(z)=QL^{(i)}_{I}(z)Q$ are respectively the
coherent Liouville operator and the interaction Liouville operator
in which $L^{(i)}_{I}(z)$ is defined by 
$L^{(i)}_{I}(z)A = 
[U\delta n_{i\uparrow}\delta n_{i\downarrow} -
\sum_{\sigma}\tilde{\Sigma}(z) n_{i\sigma}, A]$.
Operator $Q$ ($\overline{Q}$) denotes a projection operator which removes
the original operator space $\{ |a^{\dagger}_{i\sigma}) \}$
(the atomic operator space $\{ |A^{\dagger}_{i\sigma}) \}$ ).   
Note that the above expression (\ref{rself0}) is exact, 
and $\overline{G}^{(i)}_{0}(z)$ 
denotes a screened memory function in which the dynamics for the strong
atomic excitations have been removed.

We consider here the lowest-order approximation (RPT-0)~\cite{kake04-1} \!; 
we neglect in
$\overline{G}^{(i)}_{0}(z)$ a `weak' interaction Liouvillean 
$\overline{L}^{(i)}_{I}(z)\overline{Q}$.
Expanding the operator $A^{\dagger}_{i\sigma}$ by means of the nonlocal
operators $\{ a^{\dagger}_{k\sigma}
\delta(a^{\dagger}_{k^{\prime}-\sigma}a_{k^{\prime\prime}-\sigma})\}$
as $\sum_{k k^{\prime}k^{\prime\prime}}a^{\dagger}_{k\sigma}
\delta(a^{\dagger}_{k^{\prime}-\sigma}a_{k^{\prime\prime}-\sigma})
\langle k|i\rangle \langle k^{\prime}|i\rangle
\langle i|k^{\prime\prime}\rangle$,
we reach a simple form of the screened memory function
as follows (see eq. (74) in Ref. 8).
\begin{eqnarray}
\overline{G}^{(i)}_{0}(z) = 
\int \frac{d\epsilon d\epsilon^{\prime} d\epsilon^{\prime\prime}
\rho(\epsilon)\rho(\epsilon^{\prime})\rho(\epsilon^{\prime\prime})
X(\epsilon,\epsilon^{\prime},\epsilon^{\prime\prime})}
{z - \tilde{\Sigma}(z) - 
\epsilon - \epsilon^{\prime} + \epsilon^{\prime\prime}} \ .
\label{g00}
\end{eqnarray}
Here we have omitted the spin dependence for simplicity.
The correlation function 
$X(\epsilon,\epsilon^{\prime},\epsilon^{\prime\prime})$
is given by 
\begin{eqnarray}
X(\epsilon_{k},\epsilon_{k^{\prime}},
\epsilon_{k^{\prime\prime}}) & = &
\sum_{k^{\prime}_{1}k^{\prime\prime}_{1}} 
e^{i(k^{\prime}-k^{\prime\prime}-k^{\prime}_{1}+k^{\prime\prime}_{1})
\cdot R_{i}}
\langle 
\delta(a^{\dagger}_{k^{\prime}-\sigma}a_{k^{\prime\prime}-\sigma})
\delta(a^{\dagger}_{k^{\prime\prime}_{1}-\sigma}
a_{k^{\prime}_{1}-\sigma}) \rangle  \nonumber \\
& & + \sum_{k_{1}k^{\prime\prime}_{1}} 
e^{i(k-k^{\prime\prime}-k_{1}+k^{\prime\prime}_{1})
\cdot R_{i}}
\langle 
a^{\dagger}_{k^{\prime\prime}_{1}-\sigma}a_{k^{\prime\prime}-\sigma}
a_{k_{1}\sigma}
a^{\dagger}_{k\sigma} \rangle  \nonumber \\
& & - \sum_{k_{1}k^{\prime}_{1}} 
e^{i(k+k^{\prime}-k_{1}-k^{\prime}_{1})
\cdot R_{i}}
\langle 
a^{\dagger}_{k^{\prime}-\sigma}a_{k^{\prime}_{1}-\sigma}
a_{k_{1}\sigma}
a^{\dagger}_{k\sigma} \rangle \ .
\label{x}
\end{eqnarray}
The function $X$ should depend on the momentum only via
$\epsilon_{k}$, an one-electron eigenvalue in the Hartree-Fock
approximation.

The coherent potential $\tilde{\Sigma}(z)$ is obtained from a
self-consistent condition ({\it i.e.}, the CPA 
equation~\cite{elliott74,shiba71});
\begin{eqnarray}
G^{(i)}(z) = F(z) \ .
\label{cpa}
\end{eqnarray}
Note that eq. (\ref{g00}) reduces to the second order self-energy
when $X(\epsilon,\epsilon^{\prime},\epsilon^{\prime\prime})$ is 
treated by the Hartree-Fock approximation.  Thus, the
self-energy (\ref{rself0}) yields the exact
weak Coulomb interaction limit.  The self-energy 
(\ref{rself0}) also becomes exact in the atomic limit.

In order to obtain the explicit expression for the self-energy from 
eq. (\ref{x}), 
we adopted in our previous paper \cite{kake04-1} the
Hartree-Fock wave function.  
We adopt here the new wavefunction (\ref{mla}) and the
single-site approximation ({\it i.e.}, $R=0$ approximation 
\cite{kajz78}).  We have calculated the function 
$X(\epsilon_{k},\epsilon_{k^{\prime}},
\epsilon_{k^{\prime\prime}})$ in eq. (\ref{g00}).  
Actual expressions used in the
numerical calculations are given in Appendix C.  
The RPT-0 memory function obtained from the $R=0$ approximation 
in general does not satisfy the Fermi liquid condition.  Therefore, 
the lowest order approximation is not applicable for the metallic state.
Here we limit ourselves in the insulating state to demonstrate the
quantitative aspect of the MLA within the
lowest order approximation (RPT-0).
%
%
\begin{figure}
\includegraphics{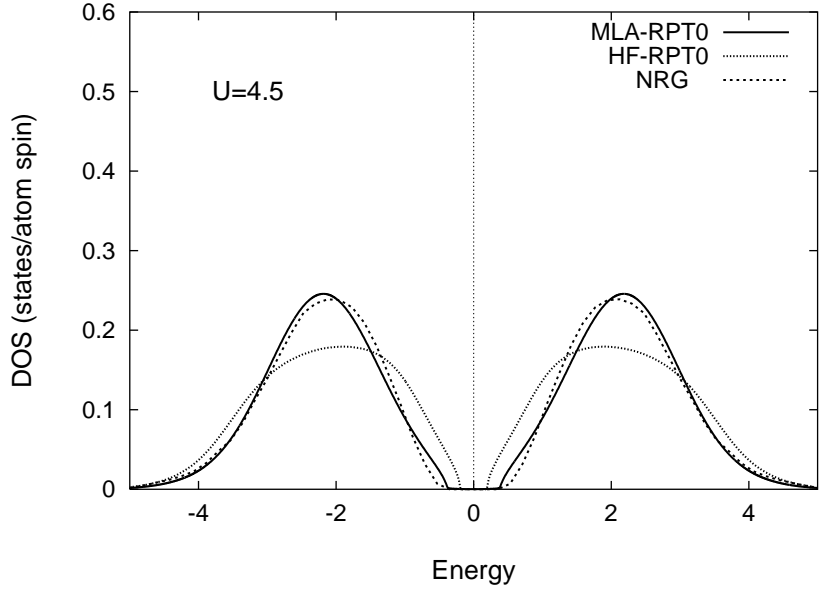}%
\caption{\label{figdos}
Single-particle excitation spectra calculated by means of the RPT-0 plus 
MLA (MLA-RPT0: solid curve), the RPT-0 plus Hartree-Fock approximation
(HF-RPT0: dotted curve)~\cite{kake04-1}, and the numerical renormalization
 group method (NRG: dashed curve)~\cite{bulla99}.
}
\end{figure}
%
%

Calculated excitation spectra in the insulating regime are presented in
Fig. \ref{figdos}.   
The result of the MLA is compared with the Hartree-Fock one and the 
NRG which is considered to be the best at zero temperature.
The spectrum with use of the Hartree-Fock wavefunction indicates the
insulator.  But the upper and lower band widths are broader than those of
the NRG.  Static correlations localize the electrons and suppress such
band broadening.
Resulting spectrum in the MLA reproduces well the NRG one.
The agreement of the spectra implies a quantitative description
of the site-diagonal Green function according to the Lehmann
representation of the Green function $G^{(i)}(z)$.  In infinite
dimensions, this means that the self-energy $\tilde{\Sigma}(z)$ is described
quantitatively by means of the present theory because of
the CPA equation $G^{(i)}(z)=F(z)$ and eq. (\ref{cohg2}).

The critical Coulomb interaction $U_{\rm c1}$ 
for gap formation is obtained from a condition that the insulator solution 
Im$\tilde{\Sigma}(0^{+})=-\infty$ disappears.  This is equivalent to the
following condition in the RPT-0 (see eq. (86) in Ref. 8).
\begin{eqnarray}
U = 4 \sqrt{c_{2}} \ .
\label{uc1-0}
\end{eqnarray}
Here $c_{2}$ is the second moment of the memory function: 
$c_{2} = \int d\epsilon d\epsilon^{\prime} d\epsilon^{\prime\prime} \,
\rho(\epsilon) \, \rho(\epsilon^{\prime}) \, 
\rho(\epsilon^{\prime\prime}) \,
(\epsilon + \epsilon^{\prime} - \epsilon^{\prime\prime})^{2} 
X(\epsilon,\epsilon^{\prime},
\epsilon^{\prime\prime})$.

For the wavefunction (\ref{mla}), the second moment $c_{2}$ is 
given by the Hartree-Fock contribution
$c^{(0)}_{2}=3/8 + 3\alpha^{2}/2$ and 
the correlation correction as  
\begin{eqnarray}
c_{2} = c^{(0)}_{2} + \frac{\displaystyle c^{(2)}_{2}}
{\displaystyle 1 
+ \langle \tilde{O}^{\dagger}_{i}\tilde{O}_{i} \rangle_{0}} \ . 
\label{c2-1}
\end{eqnarray}
Here $\alpha=1/\sqrt{\pi}$ for the hypercubic lattice and
$\alpha=4\sqrt{2}/3\pi$ for the Bethe lattice. 
The correlation contribution $c^{(2)}_{2}$ in the second
term at the r.h.s. of eq. (\ref{c2-1}) is given as follows.
\begin{eqnarray}
c^{(2)}_{2} & = & 6 \int d\epsilon_{1} d\epsilon_{2} 
d\epsilon_{3}
\rho(\epsilon_{1})\rho(\epsilon_{2})\rho(\epsilon_{3})
(\epsilon_{1}-\epsilon_{2}-\epsilon_{3})^{2}
\big[
f(\epsilon_{1})f(\epsilon_{2})\kappa_{1}(-\epsilon_{1}-\epsilon_{2})
\nonumber \\
& & \hspace{25mm}
-f(\epsilon_{1})(1-f(\epsilon_{2}))\kappa_{2}(\epsilon_{2},\epsilon_{1})
+f(\epsilon_{1})(f(\epsilon_{3})-f(\epsilon_{2}))\lambda(-\epsilon_{1})
\big] \ , \hspace{10mm}
\label{c22}
\end{eqnarray}
\begin{eqnarray}
\kappa_{1}(\epsilon_{k}) = \int d\epsilon d\epsilon^{\prime}
\rho(\epsilon)\rho(\epsilon^{\prime})f(\epsilon)(1-f(\epsilon^{\prime}))
\kappa_{0}(\epsilon_{k},\epsilon^{\prime}-\epsilon) \ ,
\label{kappa1}
\end{eqnarray}
\begin{eqnarray}
\kappa_{2}(\epsilon_{k}, \epsilon_{k^{\prime}}) = 
\int d\epsilon d\epsilon^{\prime}
\rho(\epsilon)\rho(\epsilon^{\prime})f(\epsilon)(1-f(\epsilon^{\prime}))
\kappa_{0}(\epsilon_{k}-\epsilon,\epsilon^{\prime}-\epsilon_{k^{\prime}})
\ ,
\label{kappa2}
\end{eqnarray}
\begin{eqnarray}
\lambda(\epsilon_{k}) = \int d\epsilon 
\rho(\epsilon)(1-f(\epsilon))
\lambda_{0}(\epsilon_{k}, \epsilon)
\ ,
\label{lambda}
\end{eqnarray}
\begin{eqnarray}
\kappa_{0}(\epsilon_{k},\epsilon_{k^{\prime}}) = 
\int d\epsilon d\epsilon^{\prime}
\rho(\epsilon)\rho(\epsilon^{\prime})f(\epsilon)(1-f(\epsilon^{\prime}))
\eta(\epsilon^{\prime}-\epsilon+\epsilon_{k})
\eta(\epsilon^{\prime}-\epsilon+\epsilon_{k^{\prime}}) \ ,
\label{kappa0}
\end{eqnarray}
\begin{eqnarray}
\lambda_{0}(\epsilon_{k},\epsilon_{k^{\prime}}) & = &
\int d\epsilon d\epsilon^{\prime} d\epsilon^{\prime\prime}
\rho(\epsilon)\rho(\epsilon^{\prime})\rho(\epsilon^{\prime\prime})
f(\epsilon)(1-f(\epsilon^{\prime}))f(\epsilon^{\prime\prime})
\nonumber \\
& & \hspace{20mm} \times
\eta(\epsilon^{\prime}-\epsilon-\epsilon^{\prime\prime}+\epsilon_{k})
\eta(\epsilon^{\prime}-\epsilon-\epsilon^{\prime\prime}
+\epsilon_{k^{\prime}}) \ . \hspace{16mm}
\label{lambda0}
\end{eqnarray}
Here $\eta(\epsilon_{k}) \equiv U/(\epsilon_{k}-\epsilon_{\rm c})$.
See Appendix C for the actual expression of $c^{(2)}_{2}$.

We have determined the critical Coulomb interaction $U_{\rm c1}$
solving eq. (\ref{uc1-0}).  
For a hypercubic lattice, we obtained $U_{\rm c1}(\rm MLA) = 3.237$, 
while we find
$U_{\rm c1}({\rm HF})=3.693$ when we adopt the Hartree-Fock wave function.  
The reduction of $U_{\rm c1}$ due to electron correlations is understood
from the DOS in Fig. \ref{figdos}.  There we observe that 
the electron correlations on the static matrix elements enhance the
Mott-Hubbard peaks at $\omega = \pm U/2$.
It reduces the amplitude of DOS near the Fermi level, and 
therefore lowers the critical value $U_{\rm c1}$.
%
%
\begin{table}[tb]
\caption{
Critical Coulomb interaction $U_{\rm c1}$ for the Bethe lattice in various
 approximations. MLA-RPT0: The MLA plus the lowest-order RPT in the PM-CPA 
(Present result), IPT: Iterative perturbation theory 
\cite{georges93}, ED: Exact diagonalization method \cite{caff94}, 
PSCT: Projective self-consistent technique \cite{moe95}, 
LMA: Local moment approach \cite{logan97},
NRG: Numerical renormalization group
 approach \cite{bulla99}, $1/U$ Exp.: $1/U$ expansion method 
\cite{kali02}.
}
\vspace*{5mm}
\label{table1}
\begin{tabular}{ccccccc}
\hline
MLA-RPT0 & \ \ \ IPT \ \ \ & \ \ \ ED \ \ \ & \ PSCT \ & \ \ \ LMA \ \ \ & \ \ NRG
 \ \ & \ $1/U$ Exp. \  \\
\hline
3.36 & 3.67 & 3.04 & 3.39 & 3.41 & 3.54 & 2.97 \\
\hline
\end{tabular}
\vspace*{5mm}
\end{table}
%
%

The present value $U_{\rm c1}(\rm MLA)=3.237$ quantitatively agrees 
with the numerical 
result $U_{\rm c1}(\rm NRG)=3.25$ obtained by the NRG \cite{bulla99}.
In the case of the Bethe lattice for which the noninteracting DOS is
given by $\rho(\epsilon)=\pi^{-1}\sqrt{2-\epsilon^{2}}$, 
we obtained $U_{\rm c1}(\rm MLA)=3.359$ and
$U_{\rm c1}({\rm HF})=3.827$.  
The critical value for the Bethe lattice has been
calculated by various methods 
\cite{bulla99,georges93,caff94,moe95,logan97,kali02,hub64}.  
These results are summarized in Table I
together with the present result (MLA+RPT-0).  
There are some discrepancies in $U_{\rm c1}$ among the theories in case of 
the Bethe lattice.  
Our result agrees well with $U_{\rm c1}=3.39$ obtained by the projective 
self-consistent technique (PSCT) \cite{moe95} which is exact in the low 
energy region, and $U_{\rm c1}=3.41$ obtained by the local moment 
approach (LMA) \cite{logan97}.  
Note that the results of the ED (Exact Diagonalization Method)~\cite{caff94} 
and NRG~\cite{bulla99} methods considerably depend on 
the way of line-broadening for the $\delta$-function spectrum in the
case of the Bethe lattice.
The present approach describes $U_{\rm c1}$ within
1\% error for both the hypercubic and Bethe lattices.

The quantitative description of the excitation spectra 
for the Mott insulator
with use of the RPT-0 may come as a surprise because the RPT is
an approach starting from the weak Coulomb interaction limit 
due to the expansion
of the screened memory function with respect to the interaction
Liouvillean~\cite{kake04-1}.
In this respect, it is worth pointing out that
the RPT-0 can also describe the dynamics of 
the strongly correlated electrons 
via the effective medium and atomic self-energy in the
denominator in eq. (\ref{rself0}) since the RPT-0 memory function
reduces to the Hubbard III approximation~\cite{hub64} 
in this region.

\section{Summary}
We have proposed a new local-ansatz wavefunction with momentum-dependent
variational parameters (MLA) to improve the LA by Stollhoff and 
Fulde~\cite{stoll80}.
It is constructed by using the `flexible' local operators which produce 
the two-particle excited states in the momentum space from the
Hartree-Fock state and project those states onto the local excited
states in the real space.  The best wavefunction is chosen by
controlling the momentum dependent variational parameters of the excited 
states in the momentum space on the basis of the variational principle.
We obtained the ground-state energy of the MLA 
within a single-site approximation.
Minimizing the energy, we derived a self-consistent equation for the
variational parameters, and obtained an approximate
solution which interpolates between the weak Coulomb interaction limit
and the atomic limit.  The correlation energy in the MLA agrees with the
result of the second-order perturbation theory in infinite dimensions in
the weak Coulomb interaction limit and yields the correct atomic limit
as it should be.

We have investigated numerically the validity of the theory using the
half-filled band Hubbard model in infinite dimensions.  We verified that
the MLA improves the LA in the whole range of the Coulomb interaction
energy parameter $U$.  For the hypercubic lattice, we found
that the MLA yields the correlation energy lower than that of the GA in
the range $0 < U < 3.28$.  The double occupation number in the MLA is
smaller than that of the LA irrespective of $U$.  The GA overestimates
the double occupancy in the range $0 < U \lesssim 2$, and underestimate
it in the range $U \gtrsim 3$.  We found that the MLA shows a reasonable energy
dependence of the momentum distribution in the range $0 < U \lesssim
3.0$.  This is qualitatively different from the LA and the GA because
both of them lead to the energy-independent momentum distributions below and
above the Fermi level.  The quasiparticle weights in the MLA are very
close to those of the NRG in the range $0 < U \lesssim 2.5$, while the
LA and the GA overestimate them in general.
These results suggest that the MLA is applicable to the systems with
$U/W \lesssim 1.5$, for example, the systems like transition metals and 
alloys.  Here $W$ denotes the band width of the noninteracting system.

The critical Coulomb interaction $U_{\rm c2}$ in the MLA was obtained
from the vanishment of the quasiparticle weights as $U_{\rm c2}=3.21$.  
It is comparable to $U_{\rm c2}=4.10$ in the NRG, while the LA and 
the GA give larger values $U_{\rm c2}=7.82$ and $4.51$, respectively.

We have also shown that the MLA combined with the PM-CPA is useful for
understanding the correlation effects on the excitation spectra
in the insulator regime.
The MLA wavefunction allows us to calculate the static correlations in
the retarded Green function obtained by the PM-CPA.  We have
demonstrated that the MLA+PM-CPA can quantitatively describe the
excitation spectra in the insulator regime.  Calculated critical Coulomb
interactions $U_{\rm c1}$ for a gap formation agree with the best results 
obtained by the other methods within 1 \% error for both the hypercubic 
and the Bethe lattices. 

Although the present approach interpolates between the weak Coulomb
interaction limit and the atomic limit, and much improves the LA, it does
not describe the metal-insulator transition in a self-consistent way.
The wavefunction
continuously changes from the Hartree-Fock metallic state to the atomic
one in the present theory, and does not show any anomaly at $U_{\rm c2}$
($U_{\rm c1}$) obtained from the momentum distribution (the excitation
spectra with use of the MLA+PM-CPA).  
Further improvements of the theory toward the strongly correlated region
are desired to describe the metal-insulator transition 
in a self-consistent way.
\vspace{7mm}

\section*{Acknowledgment}
The authors would like to express our sincere thanks to Prof. P. Fulde 
for valuable discussions on our wavefunction.
\vspace{3mm}

\appendix
\section{Average $\langle \tilde{A} \, \rangle$ in the single-site
 approximation} 
In this Appendix, we derive the formula (\ref{ava}) in the single-site
approximation.  Let us consider $A_{N}$ and $B_{N}$ such that 
\begin{eqnarray}
A_{N} = \Big\langle \Big[\prod_{i} (1-\tilde{O}^{\dagger}_{i}) \Big] 
\tilde{A}
\Big[ \prod_{i} (1-\tilde{O}_{i}) \Big] \Big\rangle_{0}
\ ,
\label{an}
\end{eqnarray}
\begin{eqnarray}
B_{N} = \Big\langle \Big[\prod_{i} (1-\tilde{O}^{\dagger}_{i}) \Big]
\Big[ \prod_{i} (1-\tilde{O}_{i}) \Big] \Big\rangle_{0}
\ .
\label{bn}
\end{eqnarray}
Expanding $B_{N}$ with respect to site 1, we obtain
\begin{eqnarray}
B_{N} & = & B^{(1)}_{N-1} 
- \Big\langle \tilde{O}^{\dagger}_{1}
\Big[{\prod_{i}}^{(1)} (1-\tilde{O}^{\dagger}_{i}) \Big]
\Big[ {\prod_{i}}^{(1)} (1-\tilde{O}_{i}) \Big] \Big\rangle_{0}
\nonumber \\
& &  \hspace{-15mm}
- \Big\langle
\Big[{\prod_{i}}^{(1)} (1-\tilde{O}^{\dagger}_{i}) \Big] \tilde{O}_{1}
\Big[ {\prod_{i}}^{(1)} (1-\tilde{O}_{i}) \Big] \Big\rangle_{0}
+ \Big\langle \tilde{O}^{\dagger}_{1}
\Big[{\prod_{i}}^{(1)} (1-\tilde{O}^{\dagger}_{i}) \Big] \tilde{O}_{1}
\Big[ {\prod_{i}}^{(1)} (1-\tilde{O}_{i}) \Big] \Big\rangle_{0}
\ ,  \hspace{10mm}
\label{bn1}
\end{eqnarray}
and
\begin{eqnarray}
B^{(1)}_{N-1} = \Big\langle \Big[ {\prod_{i}}^{(1)} 
(1-\tilde{O}^{\dagger}_{i}) \Big]
\Big[ {\prod_{i}}^{(1)} (1-\tilde{O}_{i}) \Big] \Big\rangle_{0} 
\ .
\label{bn-1}
\end{eqnarray}
Here the product ${\prod_{i}}^{(1)}$ means the product with respect to
all the sites except site 1.

When we calculate $B_{N}$ applying Wick's theorem, 
we neglect the contractions between
different sites.  This is a single-site approximation and then
eq. (\ref{bn1}) is expressed as
\begin{eqnarray}
B_{N} & = & \big\langle 
\big( 1 - \tilde{O}^{\dagger}_{1} \big) 
\big( 1 - \tilde{O}_{1} \big) \big\rangle_{0} \, 
B^{(1)}_{N-1}
\ .
\label{bnssa}
\end{eqnarray}
We can make the same calculations for $A_{N}$.  
In this case, there are
two-types of terms, the terms in which the operator $\tilde{O}_{1}$ is
contracted to $\tilde{A}$ and the other terms with 
$\tilde{O}_{1}$ contracted to
the operators $\tilde{O}_{i}$ $(i \ne 1)$.  We have then 
\begin{eqnarray}
A_{N} & = & \big\langle 
\big( 1 - \tilde{O}^{\dagger}_{1} \big) \tilde{A}  
\big( 1 - \tilde{O}_{1} \big) \big\rangle_{0} \, 
B^{(1)}_{N-1}
+ \big\langle 
\big( 1 - \tilde{O}^{\dagger}_{1} \big) 
\big( 1 - \tilde{O}_{1} \big) \big\rangle_{0} \, 
A^{(1)}_{N-1}
\ ,
\label{an1}
\end{eqnarray}
and
\begin{eqnarray}
A^{(1)}_{N-1} = \Big\langle \Big[ {\prod_{i}}^{(1)} 
(1-\tilde{O}^{\dagger}_{i}) \Big] \tilde{A}
\Big[ {\prod_{i}}^{(1)} (1-\tilde{O}_{i}) \Big] \Big\rangle_{0}
\ .
\label{an-1}
\end{eqnarray}
Successive application of the recursive relations (\ref{bnssa}) and
(\ref{an1}) leads to  
\begin{eqnarray}
A_{N} & = & \sum_{i} \big\langle 
\big( 1 - \tilde{O}^{\dagger}_{i} \big) \tilde{A}  
\big( 1 - \tilde{O}_{i} \big) \big\rangle_{0} \, 
B^{(i)}_{N-1}
\ ,
\label{an2}
\end{eqnarray}
\begin{eqnarray}
B_{N} = \big\langle 
\big( 1 - \tilde{O}^{\dagger}_{i} \big)  
\big( 1 - \tilde{O}_{i} \big) \big\rangle_{0} \, 
B^{(i)}_{N-1} = \prod_{i} \big\langle 
\big( 1 - \tilde{O}^{\dagger}_{i} \big)  
\big( 1 - \tilde{O}_{i} \big) \big\rangle_{0}
\ .
\label{bn2}
\end{eqnarray}
Taking the ratio $A_{N}/B_{N}$, we reach eq. (\ref{ava}).
\begin{eqnarray}
\langle \tilde{A} \rangle =  \sum_{i} \dfrac{
\big\langle 
\big( 1 - \tilde{O}^{\dagger}_{i} \big) \tilde{A}  
\big( 1 - \tilde{O}_{i} \big) \big\rangle_{0}
}
{
\big\langle \big( 1 - \tilde{O}^{\dagger}_{i} \big)  
\big( 1 - \tilde{O}_{i} \big) \big\rangle_{0}
}
\ .
\label{avassa}
\end{eqnarray}

\section{Laplace transform for the correlation
 calculations} 
Using the relation
\begin{eqnarray}
\dfrac{1}{z - \epsilon_{4} + \epsilon_{3} - \epsilon_{2} + \epsilon_{1}
 + \epsilon_{\rm c}} = -i \int^{\infty}_{0} dt \,
{\rm
e}^{i(z-\epsilon_{4}+\epsilon_{3}-\epsilon_{2}+\epsilon_{1}
+\epsilon_{\rm c}) \, t}
\ ,
\label{laplace}
\end{eqnarray}
we can reduce the number of integrals in the physical quantities.
Here $z=\omega+i\delta$, and $\delta$ is an infinitesimal positive
number.
Laplace transforms of various elements in the physical quantities are
summarized as follows.
\begin{eqnarray}
\langle \tilde{H} \tilde{O}_{i}\rangle_{0} 
& =  & \langle \tilde{O}^{\dagger}_{i}\tilde{H} \rangle_{0}^{\ast}
\nonumber \\
& = & i U^{2} [1 - \eta (1 - 2 \langle n_{i\uparrow} \rangle_{0}) 
(1 - 2 \langle n_{i\downarrow} \rangle_{0})]
\int^{\infty}_{0} \! dt \, {\rm e}^{i\epsilon_{\rm c}t} \, 
a_{\uparrow}(-t)a_{\downarrow}(-t)b_{\uparrow}(t)b_{\downarrow}(t) \ .
\hspace{5mm}
\label{lho}
\end{eqnarray}
\begin{eqnarray}
\langle \tilde{O}^{\dagger}_{i} \tilde{H} \tilde{O}_{i} \rangle_{0} & = &
- U^{2} [1 - \eta (1 - 2 \langle n_{i\uparrow} \rangle_{0}) 
(1 - 2 \langle n_{i\downarrow} \rangle_{0})]^{2}   \nonumber \\
& & \hspace{0mm} \times
\int^{\infty}_{0} \! dtdt^{\prime} 
{\rm e}^{i\epsilon_{\rm c}(t+t^{\prime})}
\big[
a_{\uparrow}(-t-t^{\prime})b_{\uparrow}(t+t^{\prime})
a_{\downarrow}(-t-t^{\prime})b_{1\downarrow}(t+t^{\prime})  \nonumber \\
& & \hspace{30mm}
-a_{\uparrow}(-t-t^{\prime})b_{\uparrow}(t+t^{\prime})
a_{1\downarrow}(-t-t^{\prime})b_{\downarrow}(t+t^{\prime})  \nonumber \\
& & \hspace{30mm}
+a_{\uparrow}(-t-t^{\prime})b_{1\uparrow}(t+t^{\prime})
a_{\downarrow}(-t-t^{\prime})b_{\downarrow}(t+t^{\prime})  \nonumber \\
& & \hspace{30mm}
-a_{1\uparrow}(-t-t^{\prime})b_{\uparrow}(t+t^{\prime})
a_{\downarrow}(-t-t^{\prime})b_{\downarrow}(t+t^{\prime})
\big] \ , \hspace{10mm}
\label{loh0o}
\end{eqnarray}
\begin{eqnarray}
\langle \tilde{O}^{\dagger}_{i} O_{i} \tilde{O}_{i} \rangle_{0} & = &
- U^{2} [1 - \eta (1 - 2 \langle n_{i\uparrow} \rangle_{0}) 
(1 - 2 \langle n_{i\downarrow} \rangle_{0})]^{2}   \nonumber \\
& & \hspace{0mm} \times
\int^{\infty}_{0} \! dtdt^{\prime} 
{\rm e}^{i\epsilon_{\rm c}(t+t^{\prime})}
\big[
a_{\uparrow}(-t)b_{\uparrow}(t+t^{\prime})
a_{\downarrow}(-t)b_{\downarrow}(t+t^{\prime})
a_{\uparrow}(-t^{\prime})a_{\downarrow}(-t^{\prime})  \nonumber \\
& & \hspace{35mm}
-a_{\uparrow}(-t)b_{\uparrow}(t+t^{\prime})
a_{\downarrow}(-t-t^{\prime})b_{\downarrow}(t)
a_{\uparrow}(-t^{\prime})b_{\downarrow}(t^{\prime})  \nonumber \\
& & \hspace{35mm}
-a_{\uparrow}(-t-t^{\prime})b_{\uparrow}(t)
a_{\downarrow}(-t)b_{\downarrow}(t+t^{\prime})
b_{\uparrow}(t^{\prime})a_{\downarrow}(-t^{\prime})  \nonumber \\
& & \hspace{38mm}
+a_{\uparrow}(-t-t^{\prime})b_{\uparrow}(t)
a_{\downarrow}(-t-t^{\prime})b_{\downarrow}(t)
b_{\uparrow}(t^{\prime})b_{\downarrow}(t^{\prime})
\big] \ , \hspace{10mm}
\label{looo}
\end{eqnarray}
\begin{eqnarray}
\langle \tilde{O}^{\dagger}_{i}\tilde{O}_{i} \rangle_{0} & = &
- U^{2} [1 - \eta (1 - 2 \langle n_{i\uparrow} \rangle_{0}) 
(1 - 2 \langle n_{i\downarrow} \rangle_{0})]^{2}  
\hspace{0mm} \nonumber \\
& & \hspace{10mm} \times
\int^{\infty}_{0} \! dtdt^{\prime} 
{\rm e}^{i\epsilon_{\rm c}(t+t^{\prime})}
a_{\uparrow}(-t-t^{\prime})b_{\uparrow}(t+t^{\prime})
a_{\downarrow}(-t-t^{\prime})b_{\downarrow}(t+t^{\prime}) \ . 
\hspace{15mm}
\label{loo}
\end{eqnarray}
Here
\begin{eqnarray}
a_{\sigma}(t) = \int d\epsilon \rho(\epsilon)
f(\epsilon+\tilde{\epsilon}_{\sigma}) \, {\rm e}^{-i\epsilon t}
\ ,
\label{asigma}
\end{eqnarray}
\begin{eqnarray}
b_{\sigma}(t) = \int d\epsilon \rho(\epsilon)
(1-f(\epsilon+\tilde{\epsilon}_{\sigma})) \, {\rm e}^{-i\epsilon t}
\ ,
\label{bsigma}
\end{eqnarray}
\begin{eqnarray}
a_{1\sigma}(t) = \int d\epsilon \rho(\epsilon)
f(\epsilon+\tilde{\epsilon}_{\sigma}) \, \epsilon \, {\rm e}^{-i\epsilon t}
\ ,
\label{a1sigma}
\end{eqnarray}
\begin{eqnarray}
b_{1\sigma}(t) = \int d\epsilon \rho(\epsilon)
(1-f(\epsilon+\tilde{\epsilon}_{\sigma})) \, \epsilon \, {\rm e}^{-i\epsilon t}
\ ,
\label{b1sigma}
\end{eqnarray}
and $\tilde{\epsilon}_{\sigma} = \epsilon_{0} 
+ U \langle n_{i-\sigma} \rangle_{0} - \mu$ is the Hartree-Fock level
measured from the Fermi level $\mu$.  $\rho(\epsilon)$ in the above
expressions denotes the density of states for $\epsilon_{k}$, {\it i.e.},
the Fourier transform of $t_{ij}$.

Correlation contribution to the momentum distribution function
(\ref{onko2}) is given by
\begin{eqnarray}
N \langle \tilde{O}^{\dagger}_{i} \tilde{n}_{k\sigma} \tilde{O}_{i} \rangle_{0} 
& = & 
- 2 U^{2} [1 - \eta (1 - 2 \langle n_{i\uparrow} \rangle_{0}) 
(1 - 2 \langle n_{i\downarrow} \rangle_{0})]^{2}   \nonumber \\
& & \hspace{-5mm} \times
\int^{\infty}_{0} \! dtdt^{\prime} 
{\rm e}^{i\epsilon_{\rm c}(t+t^{\prime})}
a_{\sigma}(t+t^{\prime})b_{\sigma}(-t-t^{\prime}) \nonumber \\
& &  \hspace{-15mm}  \times
\big[ f(\epsilon_{k\sigma}) a_{-\sigma}(t+t^{\prime}) 
{\rm e}^{-i\epsilon_{\rm c}(t+t^{\prime})}
-(1 - f(\epsilon_{k\sigma})) a_{-\sigma}(-t-t^{\prime})
{\rm e}^{i\epsilon_{\rm c}(t+t^{\prime})}
\big]
\ .
\label{lonko}
\end{eqnarray}
Correlation contribution to the electron number (\ref{ono2}) 
which appears in the
calculation of the double occupation number is expressed as
\begin{eqnarray}
\langle \tilde{O}^{\dagger}_{i} \tilde{n}_{i\sigma} \tilde{O}_{i} \rangle_{0} 
& = & 
- U^{2} [1 - \eta (1 - 2 \langle n_{i\uparrow} \rangle_{0}) 
(1 - 2 \langle n_{i\downarrow} \rangle_{0})]^{2}   \nonumber \\
& & \hspace{0mm} \times
\int^{\infty}_{0} \! dtdt^{\prime} 
{\rm e}^{i\epsilon_{\rm c}(t+t^{\prime})}
\big[
a_{-\sigma}(-t-t^{\prime})b_{-\sigma}(t+t^{\prime})
a_{\sigma}(-t-t^{\prime})b_{\sigma}(t)b_{\sigma}(t^{\prime}) \nonumber \\
& &  \hspace{28mm}
-a_{-\sigma}(-t-t^{\prime})b_{-\sigma}(t+t^{\prime})
a_{\sigma}(-t)b_{\sigma}(t+t^{\prime})a_{\sigma}(t^{\prime})
\big] \ .  \hspace{10mm}
\label{lonio}
\end{eqnarray}
The element (\ref{otildeo2}) for the calculation of the double occupancy
is expressed as
\begin{eqnarray}
\langle \tilde{O}^{\dagger}_{i} O_{i} \rangle_{0} 
+ \langle O_{i} \tilde{O}_{i} \rangle_{0} & = &
- 2iU [1 - \eta (1 - 2 \langle n_{i\uparrow} \rangle_{0}) 
(1 - 2 \langle n_{i\downarrow} \rangle_{0})]  
\hspace{0mm} \nonumber \\
& & \hspace{20mm} \times
\int^{\infty}_{0} \! dt \,
{\rm e}^{i\epsilon_{\rm c} t}
a_{\uparrow}(-t)b_{\uparrow}(t)
a_{\downarrow}(-t)b_{\downarrow}(t) \ . 
\hspace{15mm}
\label{lotildeo}
\end{eqnarray}

\section{Expressions of $X(\epsilon, \epsilon^{\prime}, 
\epsilon^{\prime\prime})$, $\bar{G}^{(i)}_{0\sigma}(z)$, and 
$c^{(2)}_{2}$ in the MLA}
In the $R=0$ approximation, the correlation function 
$X(\epsilon, \epsilon^{\prime}, 
\epsilon^{\prime\prime})$ given by eq. (\ref{x}) 
is obtained from the formula (\ref{ava}) as 
\begin{eqnarray}
X(\epsilon, \epsilon^{\prime}, \epsilon^{\prime\prime}) = 
\chi(\epsilon, \epsilon^{\prime}, \epsilon^{\prime\prime})
-\dfrac{X_{1}(\epsilon, \epsilon^{\prime}, \epsilon^{\prime\prime})}
{1+\langle \tilde{O}^{\dagger}_{i} \tilde{O}_{i} \rangle_{0}}
+\dfrac{X_{2}(\epsilon, \epsilon^{\prime}, \epsilon^{\prime\prime})}
{1+\langle \tilde{O}^{\dagger}_{i} \tilde{O}_{i} \rangle_{0}}
\ .
\label{xr0}
\end{eqnarray}
For the half-filled band, we have 
\begin{eqnarray}
\chi(\epsilon, \epsilon^{\prime}, \epsilon^{\prime\prime}) = 
f(-\epsilon)f(-\epsilon^{\prime})f(\epsilon^{\prime\prime})
+ f(\epsilon)f(\epsilon^{\prime})f(-\epsilon^{\prime\prime})
\ ,
\label{x0r0}
\end{eqnarray}
\begin{eqnarray}
X_{1}(\epsilon, \epsilon^{\prime}, \epsilon^{\prime\prime}) & = & 
-f(\epsilon^{\prime\prime})f(-\epsilon)\nu(\epsilon-\epsilon^{\prime\prime}) 
-f(-\epsilon^{\prime\prime})f(\epsilon)\nu(\epsilon^{\prime\prime}-\epsilon)
\nonumber \\
& &
+f(-\epsilon^{\prime})f(-\epsilon)\nu(\epsilon+\epsilon^{\prime}) 
+f(\epsilon^{\prime})f(\epsilon)\nu(-\epsilon-\epsilon^{\prime})
\ ,
\label{x1r0}
\end{eqnarray}
\begin{eqnarray}
X_{2}(\epsilon, \epsilon^{\prime}, \epsilon^{\prime\prime}) & = &
f(-\epsilon^{\prime})f(\epsilon^{\prime\prime})
\kappa_{1}(\epsilon^{\prime}-\epsilon^{\prime\prime}) 
-f(\epsilon^{\prime})f(\epsilon^{\prime\prime})
\kappa_{2}(-\epsilon^{\prime},\epsilon^{\prime\prime}) 
\nonumber \\
& &
-f(-\epsilon^{\prime})f(-\epsilon^{\prime\prime})
\kappa_{2}(\epsilon^{\prime},-\epsilon^{\prime\prime}) 
+f(\epsilon^{\prime})f(-\epsilon^{\prime\prime})
\kappa_{1}(\epsilon^{\prime\prime}-\epsilon^{\prime})
\nonumber \\
& &
-f(\epsilon^{\prime\prime})f(\epsilon)
\kappa_{2}(-\epsilon^{\prime\prime},\epsilon)
+f(\epsilon^{\prime\prime})f(-\epsilon)
\kappa_{1}(\epsilon-\epsilon^{\prime\prime})
\nonumber \\
& &
+f(-\epsilon^{\prime\prime})f(\epsilon)
\kappa_{1}(\epsilon^{\prime\prime}-\epsilon)
-f(-\epsilon^{\prime\prime})f(-\epsilon)
\kappa_{2}(\epsilon^{\prime\prime},-\epsilon)
\nonumber \\
& &
+f(\epsilon^{\prime})f(\epsilon)
\kappa_{1}(-\epsilon^{\prime}-\epsilon)
-f(\epsilon^{\prime})f(-\epsilon)
\kappa_{2}(\epsilon,\epsilon^{\prime})
\nonumber \\
& &
-f(-\epsilon^{\prime})f(\epsilon)
\kappa_{2}(\epsilon^{\prime},\epsilon)
+f(-\epsilon^{\prime})f(-\epsilon)
\kappa_{1}(\epsilon^{\prime}+\epsilon)
\nonumber \\
& &
+(f(\epsilon)-f(\epsilon^{\prime\prime}))
[f(-\epsilon^{\prime})\lambda(\epsilon^{\prime})
-f(\epsilon^{\prime})\lambda(-\epsilon^{\prime})]
\nonumber \\
& &
-(f(-\epsilon)-f(\epsilon^{\prime}))
[f(\epsilon^{\prime\prime})\lambda(-\epsilon^{\prime\prime})
-f(-\epsilon^{\prime\prime})\lambda(\epsilon^{\prime\prime})]
\nonumber \\
& &
+(f(\epsilon^{\prime\prime})-f(\epsilon^{\prime}))
[f(\epsilon)\lambda(-\epsilon)
-f(-\epsilon)\lambda(\epsilon)]
\ .
\label{x2r0}
\end{eqnarray}
Here $\nu(\epsilon)$ is defined by
\begin{eqnarray}\
\nu(\epsilon_{k}) = \int d\epsilon d\epsilon^{\prime}
\rho(\epsilon)\rho(\epsilon^{\prime})f(\epsilon)f(-\epsilon^{\prime})
\eta(\epsilon^{\prime}-\epsilon+\epsilon_{k}) 
\ ,
\label{nur0}
\end{eqnarray}
$\eta(\epsilon)$ is defined by 
$\eta(\epsilon)=U/(\epsilon-\epsilon_{\rm c})$.
$\kappa_{1}(\epsilon)$, $\kappa_{2}(\epsilon, \epsilon^{\prime})$, 
and $\lambda(\epsilon)$  in $X_{2}$ are defined by
eqs. (\ref{kappa1}), (\ref{kappa2}), and (\ref{lambda}), respectively.

Substituting eq. (\ref{xr0}) into eq. (\ref{g00}) and making use of the
Laplace transform (\ref{laplace}), we obtain the expression for 
$\bar{G}^{(i)}_{0\sigma}(z)$ with use of the Laplace transform as
\begin{eqnarray}
\bar{G}^{(i)}_{0\sigma}(z) = 
M_{0}(z - \tilde{\Sigma}(z))
+\dfrac{M_{2}(z - \tilde{\Sigma}(z))}
{1+\langle \tilde{O}^{\dagger}_{i} \tilde{O}_{i} \rangle_{0}}
\ ,
\label{g00r0}
\end{eqnarray}
\begin{eqnarray}
M_{0}(z) = -i \int dt \, {\rm e}^{izt} (b(-t)^{3}+b(t)^{3}) \ ,
\label{lm0r0}
\end{eqnarray}
\begin{eqnarray}
M_{2}(z) = -3i \int dt \, {\rm e}^{izt} \phi(t) \big[
\kappa_{1}(t) - \kappa_{2}(t) + 
(b(-t)-b(t)) \lambda_{1}(t)
\big] \ ,
\label{lm2r0}
\end{eqnarray}
\begin{eqnarray}
\kappa_{1}(t) = - U^{2} \int dt^{\prime} dt^{\prime\prime} 
{\rm e}^{i\epsilon_{\rm c} 
(t^{\prime}+t^{\prime\prime})}
(b(t+t^{\prime})^{2}+b(t^{\prime}-t)^{2}) 
b(t^{\prime}+t^{\prime\prime})^{2} b(t^{\prime\prime})^{2} \ ,
\label{lkappa1}
\end{eqnarray}
\begin{eqnarray}
\kappa_{2}(t) = - 2U^{2} \int dt^{\prime} dt^{\prime\prime} 
{\rm e}^{i\epsilon_{\rm c} 
(t^{\prime}+t^{\prime\prime})}
b(t^{\prime}) b(t+t^{\prime}) b(t^{\prime\prime}) 
b(t^{\prime\prime}-t) b(t^{\prime}+t^{\prime\prime})^{2} \ ,
\label{lkappa2}
\end{eqnarray}
\begin{eqnarray}
\lambda_{1}(t) = - U^{2} \int dt^{\prime} dt^{\prime\prime} 
{\rm e}^{i\epsilon_{\rm c} 
(t^{\prime}+t^{\prime\prime})}
(b(t+t^{\prime})-b(t^{\prime}-t)) 
b(t^{\prime}+t^{\prime\prime})^{3} b(t^{\prime\prime}) \ .
\label{llambda1}
\end{eqnarray}
Here $\phi(t) = \int d\epsilon \rho(\epsilon) {\rm e}^{i\epsilon t}$,
and we used the relation $a(t)=b(-t)$ for the half-filled and symmetric
band.

In the same way, the correlation contribution $c^{(2)}_{2}$ to the
second moment of the memory function ({\it i.e.}, eq. (\ref{c22})) is
expressed as
\begin{eqnarray}
c^{(2)}_{2} = - 12 U^{2} \int dt dt^{\prime} {\rm e}^{i\epsilon_{\rm c}
(t+t^{\prime})}
b(t+t^{\prime})^{2} b_{1}(t) b(t^{\prime})
\big[
b_{1}(t) b(t^{\prime}) + 
b(t) b_{1}(t^{\prime}) - \alpha \, b(t+t^{\prime})
\big] \ .
\label{lc22}
\end{eqnarray}
Here $\alpha=1/\sqrt{\pi}$ for the hypercubic lattice and 
$\alpha=4\sqrt{2}/3\pi$ for the Bethe lattice.
$b(t)$ and $b_{1}(t)$ are defined by eqs. (\ref{bsigma}) and 
(\ref{b1sigma}), respectively.

\end{document}